\documentclass[reprint,amsmath,amssymb,aps,prb,showkeys,nofootinbib,superscriptaddress]{revtex4-2} 

\usepackage[utf8]{inputenc}
\usepackage{hyperref}
\usepackage{listings}
\usepackage{graphicx}
\usepackage{float}
\usepackage{subfigure}
\usepackage{color}
\usepackage{comment}
\usepackage{algorithm}
\usepackage{algpseudocode}
\makeatletter
\edef\ftype@algorithm{\number\ftype@algorithm}% fix revtex4-2 / float.sty clash
\makeatother
\usepackage[normalem]{ulem}

\newcommand{\red}[1]{{\color{red} {#1}}}
\newcommand{\blue}[1]{{\color{blue} {#1}}}

\begin{document}

\title{Optimal sampling strategies in event-chain Monte Carlo}

\author{James Gulliford}
\affiliation{School of Engineering, University of Warwick, Coventry CV4 7AL, United Kingdom}

\author{Gareth O. Roberts}
\affiliation{Department of Statistics, University of Warwick, Coventry CV4 7AL, United Kingdom}

\author{Michael F. Faulkner}
\affiliation{School of Engineering, University of Warwick, Coventry CV4 7AL, United Kingdom}

\date{\today}

\begin{abstract}
Event-chain Monte Carlo (ECMC) has revolutionised computational sampling over recent years, providing a powerful alternative to the molecular-dynamics (MD) and Hamiltonian Monte Carlo (HMC) algorithms.  Each method outperforms the ubiquitous random-walk Metropolis algorithm by advancing particles along deterministic trajectories, but ECMC achieves this without being constrained by Newtonian dynamics.  Recent advances exploited this dynamical freedom to induce a collective particle dynamics that relax local density variations on fast timescales.  In a foundational model of $N$ pairwise-interacting particles on the 1D torus, this led to an $\mathcal{O}(N^{3/4})$ improvement on the industry-leading computational efficiency of MD and HMC -- but with an impractically small prefactor at low/high mean particle density for general attractive/repulsive interactions.  Here we present a universal framework that generalises this high-efficiency sampling strategy to all translationally symmetric pairwise models on the torus -- creating the potential to surpass MD and HMC as the state of the art.  We also numerically elucidate the collective particle  dynamics and discuss broad impact across the physical sciences and computational statistics.
\end{abstract}

\maketitle

\section{Introduction}
\label{sec:Introduction}

Deterministic sampling has revolutionised computational science over recent decades, with the molecular-dynamics (MD)~\cite{Alder1957PhaseTransition} and Hamiltonian Monte Carlo (HMC)~\cite{Duane1987HybridMonteCarlo} algorithms making particular advances in the physical sciences and Bayesian inference. Both generate deterministic particle trajectories by simulating numerical approximations of Newtonian dynamics. The former has been applied to a diverse range of problems, from protein folding~\cite{Shaw2010Atomic} to the sampling of quantum trajectories~\cite{Car1985UnifiedApproach,Marx1996AbInitioPIMD}, while the latter has been incorporated into multi-purpose software packages such as Stan, enabling high-efficiency sampling in computational statistics~\cite{Carpenter2017Stan}.

The power of these methods lies in their avoidance of diffusive exploration of the state space, in contrast with traditional random-walk Monte Carlo methods such as the Metropolis algorithm~\cite{Metropolis1953EquationOfState,Hastings1970MonteCarloSampling}. For models of $N$ particles interacting via pairwise potentials with their nearest neighbours on the 1D torus, this results in the generation of independent samples every $\mathcal{O}(N^{9/4})$ operations, as opposed to $\mathcal{O}(N^3)$ for Metropolis~\cite{Neal2011MCMCUsingHamiltonian,Beskos2013OptimalTuning,Krauth2024HMCvsECMC}. This huge scaling improvement is viewed as the benchmark in sampling efficiency, but recent years have led to a new class of deterministic sampling algorithms with the capacity for significant improvement~\cite{Bernard2009EventChain,Peters2012RejectionFreeMonteCarlo,Michel2014GeneralizedEventChain,BouchardCote2018BPS,Bierkens2019TheZigZagProcess}. These non-reversible Markov processes have been developed in parallel in statistical physics and computational statistics. The bouncy particle sampler~\cite{BouchardCote2018BPS} and zig-zag process~\cite{Bierkens2019TheZigZagProcess} have transformed sampling methodology in Bayesian computation, while event-chain Monte Carlo (ECMC)~\cite{Bernard2009EventChain,Peters2012RejectionFreeMonteCarlo,Michel2014GeneralizedEventChain} has taken centre stage in the physical sciences. 

Rather than simulating Newtonian dynamics, ECMC is designed solely to sample from the Boltzmann distribution %$\pi(\mathbf{x}) \propto \exp (-\beta U(\mathbf{x}))$, with $U(\mathbf{x})$ the differentiable potential energy of the \emph{positions vector} $\mathbf{x} = (x_1, \dots ,x_N)^T$ and $x_i$ the position of particle $i \in \{1,\dots,N\}$.  
%\blue{$\pi(\tilde{\bf x}) \propto \exp (-\beta U(\widetilde{\bf x}))$, with $U(\widetilde{\bf x})$ the differentiable potential energy of the state vector $\widetilde{\bf x}$ (defined later).}
$\pi \propto \exp (-\beta U)$, with $U$ the differentiable (with respect to the particle positions) potential energy of the system.
%It operates by essentially inverting the Metropolis approach. Rather than proposing a random particle translation before sampling a random number $\Upsilon \sim \mathcal{U}(0, 1)$ to decide whether to accept the move, ECMC samples the random number first and then continuously advances a single \emph{active particle} at some fixed velocity $v$ until the time at which the probability of accepting a sequence of infinitesimally small Metropolis moves (directed along the trajectory) reaches $\Upsilon$~\cite{Peters2012RejectionFreeMonteCarlo}.  This defines the next event following which a new active particle is selected and advances with velocity $v$.  
It operates by continuously advancing a single \emph{active particle} at some fixed velocity $v$ until a random event occurs based on the current system configuration, following which a new active particle is selected and advances with velocity $v$.  Periodic uniform resampling of the active particle and velocity then ensures ergodicity, and variants on the algorithm allow $v$ to change at events. In addition, the joint distribution over %$\mathbf{x}$ and $v$ 
the augmented state space is symmetric with respect to velocity $v$.  For models on the torus, one can therefore break velocity symmetry (defined in Appendix~\ref{breaking_vel_symm}) while maintaining ergodicity on the original state space.  For example, one can restrict to $v \in v_0 \{ \mathbf{e}_x, \mathbf{e}_y \}$ on the 2D torus, rather than sampling $v$ from the symmetric set $v_0 \{ \pm \mathbf{e}_x, \pm \mathbf{e}_y \}$ 
($v_0 > 0$ is some fixed speed).  This prevents the diffusive behaviour of random-walk Metropolis, providing a powerful alternative approach to deterministic sampling. Indeed, ECMC has been used to great effect to characterise both the 2D melting transition~\cite{Bernard2011TwoStepMelting} and broken symmetry at the Berezinskii-Kosterlitz-Thouless phase transition~\cite{Faulkner2024SymmetryBreakingBKT} -- in analogy with the Swendsen-Wang algorithm augmenting the state space of the 2D Ising model to circumvent critical slowing down~\cite{Swendsen1987Nonuniversal}. The process has been outlined in detail in Refs~\cite{Monemvassitis2023PDMPCharacterisation,Faulkner2024SamplingAlgorithms}.

Sampling the event times amounts to simulating arrival times from a nonhomogeneous Poisson process %with intensity or \emph{event rate} $\lambda(t) = \beta\max(0, v \cdot \nabla U(\mathbf{x}(t)))$~\cite{Peters2012RejectionFreeMonteCarlo,Michel2014GeneralizedEventChain,Faulkner2024SamplingAlgorithms}.  
whose intensity is a function of %$\nabla U(\mathbf{x}(t))$ 
the gradient of $U$ with respect to the particle positions
(see Section~\ref{sec:NonFactorisedECMC})~\cite{Peters2012RejectionFreeMonteCarlo,Michel2014GeneralizedEventChain,Faulkner2024SamplingAlgorithms}.  This is challenging for pairwise models of interacting particles, as each particle pair $(i,j)$ has its own term $U_{ij}$ in the potential energy $U = \sum_{j,i>j} U_{ij}$, making it impossible to invert the equation describing the next event/arrival time (the particles are indexed from $i,j = 1$ to $N$). %To circumvent this issue, one event time can be sampled 
This can be circumvented by sampling one event time for each two-particle interaction (long-range interactions can be managed exactly in $\mathcal{O}(1)$ operations with the cell-veto method~\cite{Kapfer2016Cell-vetoMCAlgorithm,Faulkner2018AllAtomComputations}). The soonest of these defines the next event, and the corresponding vetoing particle becomes active~\cite{Michel2014GeneralizedEventChain}. This leverages the fact that the Boltzmann distribution can be written as a product over two-particle probability densities $\pi_{ij} \propto \exp (-\beta U_{ij})$ -- or factors -- and redefines the stochastic process in terms of this two-particle factorisation scheme. This leads to a higher (and therefore less desirable) mean event rate than in the non-factorised case, but makes event-time sampling tractable in a much broader class of models. Similar factorisation schemes have been combined with subsampling to render the zig-zag process tractable for large data sets in Bayesian inference~\cite{Bierkens2019TheZigZagProcess}.

The unique potential of ECMC lies in its dynamics being deterministic (and non-diffusive) but not constrained by Newton’s equations. This freedom provides an array of different sampling strategies.  One particular choice modifies standard two-particle factorisation to induce a collective particle dynamics that relaxes local density variations on fast timescales~\cite{Lei2019ECMCwithFactorFields}.  For the 1D models described above, this reduces the computational effort for generating independent samples (the reciprocal of sampling efficiency) from $\mathcal{O}(N^{5/2})$ for standard two-particle factorisation to $\mathcal{O}(N^{3/2})$.  This is an $\mathcal{O}(N^{3/4})$ improvement on MD and HMC but comes at the cost of an impractically large prefactor (due to a high mean event rate) at high/low mean particle density for repulsive/attractive interactions. The huge potential of this approach cannot therefore be realised across a broad array of important applications. 

Here we present a universal framework that generalises this high-efficiency sampling strategy to all translationally symmetric pairwise models on the torus.  This represents a general analytic approach to the sampling strategy that was demonstrated numerically for the Lennard-Jones model~\cite{Lei2019ECMCwithFactorFields} and results in an $\mathcal{O}(N^{3/2})$ computational effort (on the 1D torus) that is robust to mean particle density. We also demonstrate that the computational effort of non-factorised ECMC is $\mathcal{O}(N^3)$, as it is unable to induce efficient collective particle dynamics. This explicitly demonstrates the surprising result that -- in spite of its higher mean event rate -- even standard two-particle factorisation is more efficient than the non-factorised process in the large-$N$ limit. This contrasts with the commonly held belief that standard two-particle factorisation is chosen purely for computational convenience and paves the way towards our general analytic sampling strategy becoming the preferred sampling technique for atomistic models in the physical sciences. 

The paper is organised as follows.  In Section~\ref{sec:ECMC}, we set out non-factorised ECMC, standard two-particle factorisation and the modified sampling strategy.  We then present our novel sampling strategy in Section~\ref{sec:FactorFields}.  In Section~\ref{sec:Mixing}, we elucidate the collective particle dynamics by analysing mixing.  In Section~\ref{sec:Discussion}, we discuss the universality of our results and potential impact across the physical sciences and computational statistics.  We assume throughout that particles exist on the 1D torus $\mathbb{T}$ of length $L$ and that $U = \sum_{j,i>j}U_{ij}$ with each $U_{ij}$ a differentiable function of $x_{ij}$ (meaning we do not require uniform resampling of the active particle or velocity).  Here, $x_{ij} \in \mathbb{R}$ is the 1D \emph{spring vector} from particle $j$ to $i$, which we require due to the topology of the torus.\footnote{We separate numerical indices in subscripts with a comma, e.g. $x_{N, N-1}$, $U_{a, a+1}$.}  We define it as the total displacement of particle $i$ relative to particle $j$ since the start of the simulation, plus the initial minimal separation vector from particle $j$ to particle $i$ (periodic boundary conditions are accounted for by identifying $N+1 \equiv 1$).  We then define the $[N(N - 1) / 2]$-dimensional \emph{springs vector} $\widetilde{\mathbf{x}} := (x_{1,2}, \dots , x_{1,N}, x_{2,3}, \dots , x_{2,N}, \dots, x_{N - 1, N})^T$, as well as the position $x_i \in \mathbb{T}$ of each particle $i \in \{1,\dots,N\}$ (n.b. many models on $\mathbb{T}$ can be described solely by the positions but the springs vector is more general).  
We use $\langle \cdot \rangle$ to denote thermal averages, i.e. expectations with respect to the target distribution $\pi$.

\section{Existing sampling strategies}
\label{sec:ECMC}

\subsection{Non-factorised ECMC}
\label{sec:NonFactorisedECMC}

The most general form of \emph{non-factorised ECMC} constructs the following continuous-time Markov process.  Its state at time $t$ 
%is augmented such that it is characterised by the original 
consists of the %positions vector ${\bf x}(t) = (x_1(t), \ldots x_N(t))^T$, 
springs vector $\widetilde{\bf x}(t)$, an index $a(t) = 1, 2, \dots , N$ denoting the current active particle %(so that $a(t)$ takes one of the values $1, 2, \ldots , N$) and all other particles are stationary) 
and the velocity $v(t) = \pm v_0$ of the active particle.  The system evolves with constant $a$ and $v$ until an {\em event} occurs.  
%This event occurs according to the distribution of the first event of a non-homogeneous Poisson Process of rate $\max (0, v(t) \partial_{a(t)} U(x(t)))$.
The corresponding \emph{event time} $t_k$ is the next event or arrival time of a nonhomogeneous Poisson process with intensity or \emph{event rate} $\lambda(t) = \beta \max (0, v(t) \partial_{a(t)} U(\widetilde{\mathbf{x}}(t)))$, where $\partial_a U$ denotes the partial derivative with respect to the position $x_a$ of particle $a$.  Particle $j \in \{1, \dots, N\}$ and velocity $v = \pm v_0$ are then selected as the next active particle and velocity with probability $\propto \max(0, -v\partial_j U(\widetilde{\mathbf{x}}(t_k)))$ and the system evolves with this new $a$ and $v$ until the next event time $t_{k+1} > t_k$.  For clarity, the event times are denoted $t_1, t_2, \dots $ with $0 < t_1 < t_2 < \dots$, and $t_k$ is simulated by sampling $\Upsilon \sim \mathcal{U}(0,1)$ and then inverting 
\begin{equation}\label{ecmc_master_equation}
   - \log \Upsilon = \beta \int_{t_{k-1}}^{t_k}\max(0, v(t) \partial_{a(t)} U(\widetilde{\mathbf{x}}(t))) dt ,
\end{equation}
with $t_0 = 0$.  We note that Ref.~\cite{Peters2012RejectionFreeMonteCarlo} presented a physically intuitive construction of the event rate $\lambda(t)$ by considering a sequence of lifted Metropolis moves in a fixed direction in the continuum-time limit. 

The process of choosing the next active particle and velocity (following each event) defines the lifting scheme.  The lifting scheme described above allows $v$ to flip at each event, but we choose an alternative scheme that fixes $v$.  This is the lifting scheme described in Section~\ref{sec:Introduction} and is valid as $U$ is a function of the %separation vectors $x_i - x_j$ (i.e. it is translationally symmetric)
springs vector $\widetilde{\bf x}$ (i.e. it is symmetric to global particle translations, defining \emph{translational symmetry}) and the particles exist on the torus (see Appendix~\ref{breaking_vel_symm}).  % When implemented on $\mathbb{R}^d$ and without velocity refreshments, however, 
Indeed, employing this lifting scheme without uniform resampling of $v$ would result in a nonergodic process on $\mathbb{R}$, but breaking velocity symmetry in this way avoids diffusive behaviour and is therefore a common choice on the torus.  We set $v = v_0$ throughout and do not employ uniform resampling of either $v$ or $a$.

\subsection{Standard-factorised ECMC}
\label{sec:StandardFactorisedECMC}

It is typically challenging to solve Eq.~\eqref{ecmc_master_equation} in practice.  \emph{Standard-factorised ECMC} instead independently samples one event time $t^k_{aj}$ for each particle $j \ne a$ by inverting Eq.~\eqref{ecmc_master_equation} with $U$ and $t_k$ replaced with $U_{aj}$ and $t^k_{aj}$, respectively.  The next event time $t_k$ is then $\min(t^k_{aj})$ and the vetoing particle $w = \text{arg}\min_j t^k_{aj}$ is selected as the next active particle.  %This has the effect of defining a new stochastic process with pairwise event rates \blue{$\lambda_{ij} = \beta\max(0, \partial_a U_{ij}(x_i - x_j))$, where the dependence on $t$ is now implicit}.
This defines a new stochastic process with pairwise event rates $\lambda_{ij} = \beta v_0 \max(0, \partial_a U_{ij}(x_{ij}))$, with the $t$ dependence now implicit.

There are two significant consequences of this factorisation scheme. Firstly, the event rate tends to increase: 
\begin{align}
\label{eq:TriangleInequality}
    \beta v_0 \max \left( 0, \sum_{j \ne a} \partial_a U_{aj} \right) \le \sum_{j \ne a} \beta v_0 \max \left( 0, \partial_a U_{aj} \right) ,
\end{align} 
i.e. $\lambda \le \sum_{j \ne a} \lambda_{aj}$.
%(n.b. here we present the framework for nearest-neighbour interactions only but it generalises to all models (e.g. see Section~\ref{sec:SoftDisks})). %By the triangle inequality, $\Lambda_{\text{\rm nf}} = \beta \langle | F_{a, a-1}+F_{a, a+1}|\rangle/2 \leq \beta (\langle |F_{a, a-1}| \rangle + \langle |F_{a, a+1}| \rangle)/2 = \Lambda_{\text{f}}$, with the strict inequality $| F_{a, a-1}+F_{a, a+1}| < |F_{a, a-1}| + |F_{a, a+1}|$ holding for the overwhelming majority of configurations\footnote{Indeed, for any given particle configuration, the configurational event rate of the nonfactorised process is never greater than that of the factorised process: $\beta \max (0, -vF_a) \le \beta \max (0, -vF_{a,a-1}) + \beta \max (0, -vF_{a,a+1})$.}.  
The non-factorised dynamics therefore benefit from the cancellation of opposing forces acting on the active particle, resulting in a diminished net effective force and an increased distance travelled through the state space between events. Factorisation removes this cancellation mechanism, inflating the event rate.
%Factorisation removes this cancellation mechanism and the active particle comes to a stop sooner.  This is also reflected in the stronger condition on the configurational event rates for any given particle configuration: $\beta \max (0, -vF_a) \le \beta \max (0, -vF_{a,a-1}) + \beta \max (0, -vF_{a,a+1})$.

\begin{figure}[t]
    \includegraphics[width=\linewidth]{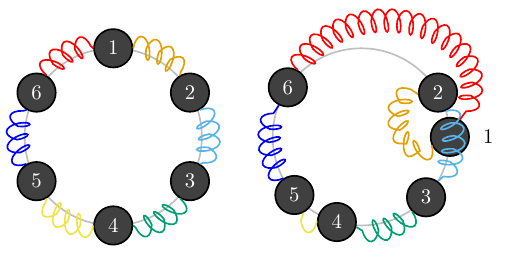}
    \caption{Schematic for both the index space $\mathcal{I}$ (defined in Section~\ref{sec:StandardFactorisedECMC}) and generalised harmonic chain (defined in Section~\ref{sec:HarmonicChain}).  The left/right subfigure is an example configuration at/perturbed from the mode.  The particle ordering $\{1, \dots, N \}$ within $\mathcal{I}$ is fixed based on the modal state-space configuration at the start of the simulation, i.e.~assuming starting at the mode (e.g.~left).  The state-space ordering at a later time may not therefore correspond to the index-space ordering (e.g.~right).  In the generalised harmonic chain, particle interactions are mediated by 1D spring vectors $x_{ij} \in \mathbb{R}$ between neighbours $(i,j)$ on $\mathcal{I}$, subject to the constraint $\sum_i x_{i+1,i} = L$.  A countably infinite number of spring configurations correspond to any given particle configuration.  The mode is given by $x_{i+1,i} = L / N$ for all $i$ (left) while spring vectors can be negative away from the mode (e.g.~the orange spring in the right subfigure).  Black circles represent particles.  Each number is the index-space value of the corresponding particle.  Curly lines represent springs, with colours denoting the same spring across subfigures.  Spring shapes have been modified for illustrative purposes and do not reflect the magnitudes $|x_{ij}|$.}
    \label{fig:HarmonicChainSchematic}
\end{figure}

The second significant consequence is modified lifting dynamics.  %, i.e. the selection of a new active particle following some event.  
In non-factorised ECMC, any non-active particle $j$ is selected as the next active particle with probability $\propto \max(0, -\partial_j U(\widetilde{\mathbf{x}}(t_k)))$ at event time $t_k$, making the particle positioned for the largest subsequent downhill movement most likely to be selected.  Whilst one of the neighbouring particles will typically be selected, non-neighbouring particles will be chosen at some finite proportion of lifting events.  This leads to $a$ making non-local jumps on the space of possible active particles, with the proportion of non-local jumps increasing as some function of $N$ (we call this space the \emph{index space} $\mathcal{I}$, defined as the set $\{1, \dots, N \}$ with toroidal topology, and fix the particle ordering $\{1, \dots, N \}$ based on the modal state-space configuration at the start of the simulation (see Fig.~\ref{fig:HarmonicChainSchematic})).  This hinders the ability of non-factorised ECMC to efficiently relax local density variations via collective motion.  This is the origin of the $\mathcal{O}(N^3)$ autocorrelation time demonstrated numerically in Section~\ref{sec:HarmonicChain}.
%We demonstrate below in \blue{Section XX} that this leads to independent samples being generated only every $\mathcal{O}(N^3)$ operations. 
%In the factorised case, lifting becomes deterministic, with the vetoing particle always becoming active following an event. 
In the standard-factorised case, the vetoing particle always becomes active following an event, making the dynamics on the index space strictly local for nearest-neighbour models.  This is accompanied by standard-factorised ECMC generating independent samples every $\mathcal{O}(N^{5/2})$ operations~\cite{Lei2019ECMCwithFactorFields}. %Whilst exhibiting superior scaling to the non-factorised case, factorisation induces large prefactors on the computational effort due to the aforementioned inflated mean event rate. 
More generally when choosing a sampling strategy, there can be a tradeoff between a low mean event rate (e.g. without factorisation) and improved lifting dynamics (e.g. with standard factorisation). 

\subsection{Factor-field ECMC}
\label{sec:FactorFieldECMC}

The $\mathcal{O}(N^{5/2})$ computational effort can be improved further still as %there is a high probability that the vetoing particle is the negative/positive neighbour (on the index space) for attractive/repulsive interactions whose strength decays with separation distance (on the state space), defining a backwards/forwards event.  
the next event is predominantly a forwards/backwards event (i.e. triggered by the forwards/backwards neighbour on the index space $\mathcal{I}$) for repulsive/attractive interactions whose strength decays with minimal separation distance.  This leads to the index of the currently active particle drifting forwards/backwards through $\mathcal{I}$, % which is suboptimal for the relaxation of high-density regions on the state space. 
which is undesirable when relaxing local density variations~\cite{Lei2019ECMCwithFactorFields}.
%This results from the combination of two-particle factorisation, broken velocity symmetry and nonzero pressure -- and leads to the drift on the index space.  
%The drift is undesirable when relaxing local density variations, leading to the suboptimal autocorrelation time~\cite{Lei2019ECMCwithFactorFields}.  
The concept of \emph{factor-fields}~\cite{Lei2019ECMCwithFactorFields,Maggs2022LargeScaleDynamicsOfECMC,Krauth2024HMCvsECMC} was introduced to generate the collective particle dynamics that relax local density variations on fast timescales.  %\blue{This is achieved by slowing exploration of the index space $\{ 1, \dots, N \}$ from ballistic to superdiffusive, where extended periods of moving in a single direction through $\{ 1, \dots, N \}$ are followed by reversals of direction \cite{Lei2019ECMCwithFactorFields}.  This provides the flexibility to select the next active particle optimally for efficient exploration of the original state space (we elucidate this numerically in Section \ref{sec:Mixing}).} 
This is achieved by superdiffusive index-space dynamics, %corresponding to 
which characterise extended periods advancing in a single direction on $\mathcal{I}$, each followed by a direction reversal~\cite{Lei2019ECMCwithFactorFields}.  This leads to slower exploration of $\mathcal{I}$ than achieved by the ballistic drift at $b=0$, but selects the next active particle optimally for efficient exploration of the original state space.  We elucidate this numerically in Section \ref{sec:Mixing}. 

For models of particles on the 1D torus of length $L$, %this amounts to 
the concept is constructed mathematically by adding the non-physical \emph{factor-field potential} $U_{\text{ff}}(b) = - b \sum_i x_{i + 1, i} = -bL$ to the potential energy, where the factor-field $b \in \mathbb{R}$ is fixed throughout. %and $x_{ji}$\footnote{We separate numerical indices in subscripts with a comma, e.g. $x_{N, N-1}$, $U_{a, a+1}$.} is the net displacement of particle $j$ relative to particle $i$ (periodic boundary conditions are accounted for by identifying $N+1 \equiv 1$, and the net displacement is the total displacement of particle $j$ from particle $i$ since the start of the simulation, plus the initial minimal separation vector from particle $i$ to particle $j$).  
Adding this constant leaves the target distribution unchanged, but, with $b$ optimally chosen (in the large-$N$ limit) to nullify the pressure $P := \beta^{-1} \partial \log Z / \partial L$, sampling an additional event for each $-b x_{i + 1, i}$ %[minimises the mean event rate by inducing] 
induces equal numbers of forwards and backwards events on long timescales, resulting in zero drift on the index space ($Z$ is the partition function and adding the factor-field potential $U_{\text{ff}}$ modifies the pressure: $P \mapsto P - b$).  This leads to the $\mathcal{O}(N^{3/2})$ autocorrelation time of factor-field ECMC~\cite{Lei2019ECMCwithFactorFields} and is analogous to introducing an auxiliary hyperparameter to minimise Bayesian model misspecification~\cite{Faulkner2024SamplingAlgorithms}. % We demonstrate this efficient collective particle dynamics in Section~\ref{sec:Mixing}.
Extensions to the higher dimensional torus achieved similarly strong performance benefits~\cite{Maggs2022LargeScaleDynamicsOfECMC}.  We emphasise that $U_{\rm ff}$ does not alter the dynamics of the Metropolis, MD or HMC algorithms~\cite{Krauth2024HMCvsECMC}.

\begin{algorithm}[t]
\caption{Sampling event $k$ with cell-horizon ECMC in the absence of singularities}\label{alg:cell_horizon}
\begin{algorithmic}
\Require %current positions vector $\mathbf{x} = \mathbf{x}(t_{k - 1})$, 
current springs vector $\widetilde{\mathbf{x}} = \widetilde{\mathbf{x}}(t_{k - 1})$, current active particle $a = a(t_{k - 1})$, %\blue{current active-particle position $x_a = x_a(t_{k - 1})$}, 
cell horizon $x_h$, factor-field $b$, inverse temperature $\beta$, number of particles $N$, speed $v_0$
\For{$j \gets 1$ \textbf{to} $N$}
\State $\widetilde{\lambda}^{\text{max}}_{aj} \leftarrow \max\{\widetilde{\lambda}_{aj}(x, b) : x_{aj} \leq x \leq x_{aj} + x_h\}$
\State $\eta^k_{aj} \sim \text{Exp}(\widetilde\lambda^{\text{max}}_{aj})$
\EndFor
\State $\eta_k \leftarrow \min(\eta^k_{aj})$
\State $w \leftarrow \text{arg}\min_j(\eta^k_{aj})$
\If{$\eta_k > x_h$}
    %\State $x_a \leftarrow x_a +  x_h$
    \State $x_{aj} \leftarrow x_{aj} + x_h $ \textbf{for all} $j \ne a$ (i.e. $x_a \leftarrow x_a + x_h $)
    %\State $a \leftarrow a$ 
    \State $t_k \leftarrow t_{k - 1} + x_h / v_0$
\Else
    %\State $x_a \leftarrow x_a + \eta_k$
    \State $x_{aj} \leftarrow x_{aj} + \eta_k$ \textbf{for all} $j \ne a$ (i.e. $x_a \leftarrow x_a + \eta_k$)
    \State $t_k \leftarrow t_{k - 1} + \eta_k / v_0$
    \State $\Gamma \sim \mathcal{U}(0,1)$
    \If{$\Gamma \leq \widetilde\lambda_{aw}(x_{aw}, b)/\widetilde\lambda^{\text{max}}_{aw}$}
        \State $a \leftarrow w$
    \EndIf
\EndIf
\end{algorithmic}
\end{algorithm}

The mean event rate at optimal $b$ is, however, even higher than the $b = 0$ case for general differentiable interaction potentials -- a problem that worsens at high/low mean particle density for repulsive/attractive models (due to a large optimal $b$).  More precisely, the mean event rate becomes linearly dependent on $b$ due to the independent sampling of the factor-field events (seen by extending the triangle inequality of Eq.~\eqref{eq:TriangleInequality}).  
This manifests itself in a large prefactor on the computational effort required to generate independent samples. %-- with the computational effort defined as the product of the mean event rate with the autocorrelation timescale (i.e. the number of operations required to generate independent samples). 
This is the key metric of sampling performance because the autocorrelation time reflects the efficiency of the dynamics on the index space, whereas the computational effort also completely encompasses the mean distance that the active particle advances on the state space between events (n.b. changes on the state space occur between events while changes on the index space occur at events).  This results in an impractically large computational effort for models with a large optimal factor-field $b$. 
%The aforementioned inflated mean event rate induces, however, large prefactors on the total computational effort required to generate independent samples.  There is therefore a tradeoff between a low mean event rate (as achieved in the non-factorised case) and improved lifting dynamics (as achieved by factorisation).
%More generally when choosing a sampling strategy, there is typically a tradeoff between a low mean event rate (as achieved in the non-factorised case) and improved lifting dynamics (as achieved by factorisation). %The respective autocorrelation scalings demonstrate that an inflated mean event rate is a worthwhile price to pay, and that factorising should not only be performed for computational tractability, but also happens to optimise sampling efficiency. 

\section{Cell-horizon ECMC}
\label{sec:FactorFields}

\begin{figure*}[t]
    \includegraphics[width=\linewidth]{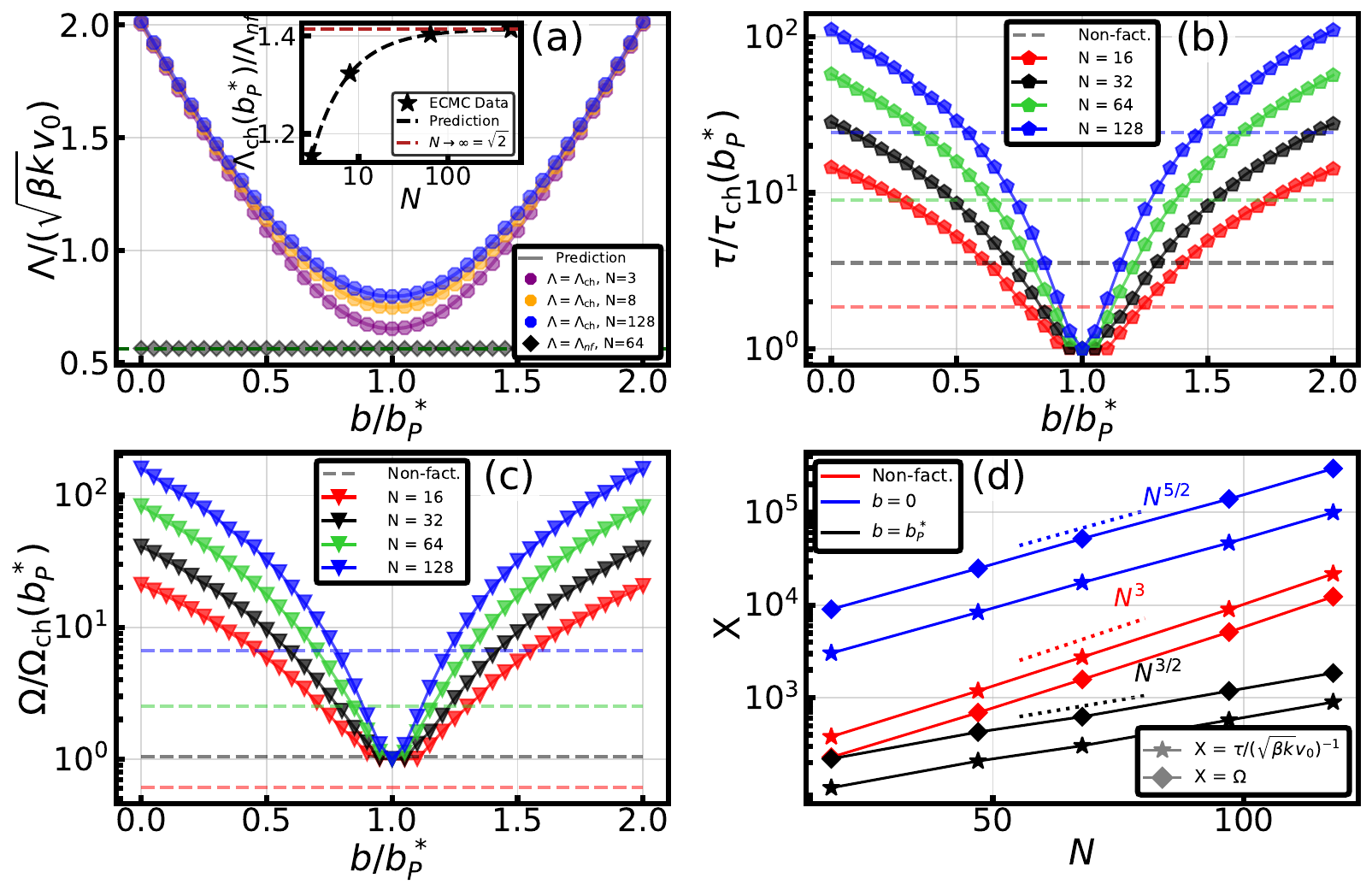}
    \caption{Comparative performance of cell-horizon ECMC against both non-factorised and standard-factorised ECMC.  Metrics are measured for the $p=2$ generalised harmonic chain with mean particle density $\rho = \sqrt{\beta k}/2$ ($(\beta k)^{-1/p}$ sets a lower length scale).  (a) MC estimates and analytic predictions of the mean event rate of cell-horizon ECMC ($\Lambda_{\rm ch}$) and non-factorised ECMC ($\Lambda_{\rm nf}$) as functions of $b$ and $N$ (n.b.~$b$ does not modify non-factorised ECMC, as predicted).  We observe excellent agreement between the estimates and predictions, with $\Lambda_{\rm ch}$ minimised at $b=b^*_P$, as predicted.  Inset: The ratio $\Lambda_{\rm ch}(b=b^*_P)/\Lambda_{\rm nf}$ emphasises that non-factorised ECMC outperforms cell-horizon ECMC for all $N > 2$ by this metric.  % (b)  MC estimates of the autocorrelation time $\tau$ of the $q=1$ structure factor $S(q) = \left| \sum_{j=1}^{N} \exp(i q x_j) \right|^2 / N$.
    (b,c)  MC estimates of the autocorrelation time $\tau$ [(b)] and computational effort $\Omega$ [(c)] of both cell-horizon ECMC and non-factorised ECMC as functions of $b$ and $N$ (both metrics are defined in Section~\ref{sec:FactorFields} and are normalised by their respective optimal cell-horizon values, $\tau_{\rm ch}(b^*_P)$ and $\Omega_{\rm ch}(b^*_P)$).  Cell-horizon results are minimised at $b=b^*_P$.  (d) MC estimates of $\tau$ and $\Omega$ for each method (at $b=b^*_P$ for cell-horizon ECMC; $b=0$ is standard-factorised ECMC) against $N$.  Results are consistent with both metrics being $\mathcal{O}(N^3)$, $\mathcal{O}(N^{5/2})$ and $\mathcal{O}(N^{3/2})$ for non-factorised, standard-factorised and optimal cell-horizon ECMC.  Optimal cell-horizon ECMC significantly outperforms the other methods by both metrics (n.b.~$\Omega$ is the key metric of sampling performance).  Lines are a guide to the eye.  Results were averaged across 10-12 independent simulations, each of between $10^5$ and $2 \times 10^6$ samples (see Appendix~\ref{sec:CodeAndSimulationData} for precise details).}
    \label{fig:HarmonicChain}
\end{figure*}

\subsection{Framework}
\label{sec:CellHorizonFramework}

\begin{comment}
For the factorised process applied to attractive/repulsive interactions, there is a high probability that the vetoing particle is the negative/positive neighbour, defining a backwards/forwards event.  This leads to a drift on the index space which is accompanied by a nonzero macroscopic pressure.  The drift is undesirable when relaxing local density variations, leading to a suboptimal autocorrelation time~\cite{Lei2019ECMCwithFactorFields}. 

\textit{factor-fields} are auxiliary parameters introduced to cancel the drift on the index space when optimally chosen to nullify the pressure.  %This is achieved by inducing equal numbers of forwards and backwards events on long timescales.  
The factor-field $b \in \mathbb{R}$ is introduced to the model by adding the factor-field potential 
\begin{equation}
\label{eq:FFpotential}
  U_{\text{ff}}(b) = - b \sum_{i = 1}^{N} x_{i + 1, i}  = -bL
\end{equation}
to the potential energy $U$.  The toroidal topology results in the constant $-bL$ on the right-hand side of Eq. \eqref{eq:FFpotential}.  This leaves the Boltzmann distribution invariant, but absorbing each $-bx_{i + 1, i}$ into its corresponding two-particle potential %can lead to significant improvements in sampling efficiency~\cite{Lei2019ECMCwithFactorFields}. 
induces equal numbers of forwards and backwards events on long timescales.
\end{comment}

Introducing the factor-field results in the \emph{total potential} $\widetilde{U} = U + U_{\rm ff}$.  For index-neighbour models (models with interactions only between index-space neighbours), this leads to the active particle experiencing a total potential 
\begin{comment}
\begin{equation}\label{prev_ff_strat}
    \begin{aligned}
    \widetilde{U}_a(x_{a,a+1}, x_{a,a-1}, b) %&= U^{\text{tot}}_{a,a-1} + U^{\text{tot}}_{a,a+1}  \\
    &= U_{a,a-1}(x_{a,a-1}) - b x_{a, a-1}  \\
    &+ U_{a,a+1}(x_{a,a+1}) - b x_{a, a+1} .
    \end{aligned}
\end{equation}
\end{comment}
\begin{equation}\label{prev_ff_strat}
    U_{a,a-1}(x_{a,a-1}) - b x_{a, a-1} + U_{a+1,a}(x_{a+1,a}) - b x_{a+1,a} .
\end{equation}
Previous non-numerical sampling strategies for implementing factor-fields for general index-neighbour models sample one event for each of the four terms %on the right-hand side of 
in Eq.~\eqref{prev_ff_strat}~\cite{Krauth2024HMCvsECMC}.  This {strategy of \emph{four-factor ECMC}} amounts to sampling two events per pairwise interaction  (i.e. one event due to the original potential and one factor-field event) and %Whilst selecting $b$ to nullify the pressure leads to optimal dynamics on the index space, the mean event rate is inflated relative to the $b=0$ case, limiting the associated reduction in computational effort. The increase in the mean event rate is linear in $b$, meaning at low/high density for attractive/repulsive interactions, where a large value of $b$ is needed to cancel the pressure, the computational effort becomes impractically large. 
leads to the impractically large computational effort described in Section \ref{sec:ECMC}.  To combat this, we propose \emph{cell-horizon ECMC} (Alg.~\ref{alg:cell_horizon}).  This novel sampling strategy samples only a single event per pairwise interaction, leading to two main benefits compared with four-factor ECMC: i) the mean event rate for optimally chosen $b$ decreases relative to $b=0$, and ii) inverting Eq. \eqref{ecmc_master_equation} (often challenging) is avoided entirely in the absence of singularities. Our approach therefore reduces the computational effort at optimal $b$ and %remains efficient at all particle densities. 
is robust to mean particle density.  

For index-neighbour models, the method writes the total potential as $\widetilde{U} = \sum_i (U_{i+1,i} - b x_{i+1,i})$ and redefines the stochastic process in terms of the extended two-particle factorisation scheme with pairwise probability densities $\widetilde{\pi}_{i+1,i} \propto \exp (-\beta (U_{i+1,i} -b x_{i+1,i}))$.  This extends the pairwise event rate to 
\begin{equation}\label{ff_event_rate}
\widetilde{\lambda}_{ij}(x_{ij}, b) \! = \! 
\begin{cases} 
\beta v_0 \! \max\left(0, \partial_a U_{ij}(x_{ij}) \! - \! b\right) \!\! & \!\! {\rm if} \,\, (i,j) \! = \! (a, a \! - \! 1) \\
\beta v_0 \! \max\left(0, \partial_a U_{ij}(x_{ij}) \! + \! b\right) \!\! & \!\! {\rm if} \,\, (i,j) \! = \! (a \! + \! 1, a) \\ 
\beta v_0 \! \max\left(0, \partial_a U_{ij}(x_{ij}) \right) & \!\! \text{otherwise} ,
\end{cases}
\end{equation}
where %$\xi = \pm 1$ for attractive/repulsive potentials and 
the third case generalises to non-index-neighbour models.  This defines the factorisation scheme of cell-horizon ECMC.  In the absence of singularities, each pairwise event time $t^k_{aj}$ is sampled from a homogeneous Poisson process with intensity $\widetilde\lambda^{\text{max}}_{aj} = \max\{\widetilde\lambda_{aj}(x, b) : x^{k - 1}_{aj} \leq x \leq x^{k - 1}_{aj} + x_h\}$, where $x_h > 0$ is the \emph{cell horizon} and %$x^{k - 1}_{ij}$ is the value of $x_{ij}$ at the previous event time $t_{k - 1}$ 
$x^{k - 1}_{ij} = x_{ij}(t_{k - 1})$ (singularities are addressed in Section~\ref{sec:SoftDisks}).  This generates corresponding event distances $\eta^k_{aj}$ and the shortest of these defines the next event, with \emph{event distance} $\eta_k = \min(\eta^k_{aj})$ and corresponding vetoing particle $w = \text{arg}\min_j(\eta^k_{aj})$ ($\eta_k$ is the distance travelled between events $k - 1$ and $k$).  If $\eta_k > x_h$, the active particle moves a distance $x_h$ to the edge of the cell and remains active (i.e. $\eta_k$ is set to $x_h$). %following the event.  
Otherwise, the active particle moves a distance $\eta_k$ and the vetoing particle becomes active with probability $\widetilde\lambda_{aw}(x^{k - 1}_{aw}+\eta_k) / \widetilde\lambda_{aw}^{\text{max}}$, defining a \emph{confirmed event}.  This process of thinning ensures convergence on the target distribution. For consistency, we set $t_k = t_{k-1} + \eta_k / v_0$ in all three cases. We note that the update of $x_{aj}$ for all $j \ne a$ in Alg.~\ref{alg:cell_horizon} (and also Alg.~\ref{alg:cell_horizon_singularities} below) is expressed in terms of each $x_{aj}$ for consistency with our formalism. As each spring vector $x_{ij}$ changes only through the displacement of particle $i$ or $j$, in practice only the position $x_a$ of the active particle is updated at each event time, i.e. each event requires only a single position update. 

Selecting $x_h$ to maximise the acceptance probability is required to optimise the computational effort. Small $x_h$ leads to a large number of rejections due to hitting the cell horizon too frequently, whereas large $x_h$ results in a %high value of $\widetilde\lambda^\text{max}$ so that the thinning acceptance probability is small. 
large $\widetilde\lambda^\text{max}$ and a low thinning acceptance rate.  The computational effort $\Omega$ is defined as the product of the mean event rate (the long-time limit of the number of confirmed events per unit time) and autocorrelation time divided by the acceptance rate, with the optimal $b$ value (denoted $b^*$) defined as that which minimises this metric. This definition reflects the number of CPU operations required to generate independent samples, assuming the cell-veto method~\cite{Kapfer2016Cell-vetoMCAlgorithm} is used for long-range interacting particles.  In such cases, we recommend incorporating the cell-veto method into cell-horizon ECMC.

\subsection{Generalised harmonic chain}
\label{sec:HarmonicChain}

\begin{figure*}[htb]
    \includegraphics[width=\linewidth]{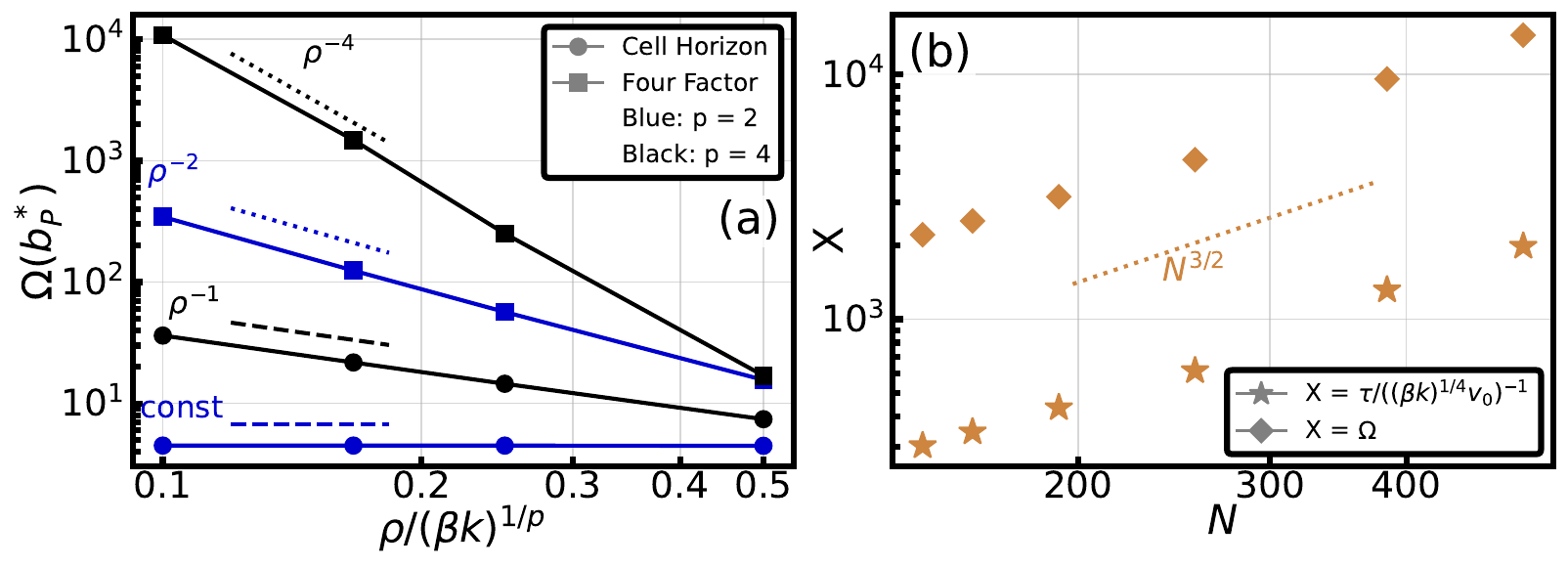}
    \caption{Comparative performance of cell-horizon ECMC against four-factor ECMC.  (a) MC estimates of the computational effort $\Omega$ (defined in Section~\ref{sec:FactorFields}) of both methods as a function of $\rho$ at fixed $N = 16$ and at optimal $b = b^*_P$ for both the $p=2$ and $p=4$ generalised harmonic chain ($(\beta k)^{-1/p}$ sets a lower length scale).  For $p=2$, results are consistent with $\Omega$ being insensitive to $\rho$ (cell-horizon ECMC) and $\mathcal{O}(\rho^{-2})$ (four-factor ECMC).  For $p=4$, results are consistent with $\Omega$ being $\mathcal{O}(\rho^{-1})$ (cell-horizon ECMC) and $\mathcal{O}(\rho^{-4})$ (four-factor ECMC).  This reflects cell-horizon ECMC circumventing the impractically large $\Omega$ of four-factor ECMC at large $b^*$ (at small $\rho$ in this case).  Lines are a guide to the eye.  (b) MC estimates of the autocorrelation time $\tau$ (defined in Section~\ref{sec:FactorFields}) and $\Omega$ of cell-horizon ECMC as a function of $N$ at fixed $\rho = (\beta k)^{1/4} / 2$ for the $p=4$ generalised harmonic chain.  Results are consistent with $\tau, \Omega = \mathcal{O}(N^{3/2})$.  Results were averaged across 10-12 independent simulations, each of between $10^5$ and $10^6$ samples (see Appendix~\ref{sec:CodeAndSimulationData} for precise details).}
    \label{fig:harm_chain_figures_2}
\end{figure*}

We explore cell-horizon ECMC with the generalised harmonic-chain model on the 1D torus of length $L$.  Particles interact via springs that mediate polynomial potentials between neighbours on the index space. 
The total potential is a function of the springs vectors $\widetilde{\bf x}$: %Including the factor-field, the total potential is 
\begin{align}
    \widetilde{U}(\widetilde{\bf x}, b) = \frac{k}{p} \sum_i |x_{i+1,i}|^p-b \sum_i x_{i+1,i} \label{hc_factor1} ,
\end{align}
where $k > 0$ is the spring stiffness, $p \ge 1$, and the springs are subject to the constraint $\sum_i x_{i+1,i}=L$.  Fig.~\ref{fig:HarmonicChainSchematic} presents schematics of the model at the mode (left) and far from equilibrium (right) (the latter is an example containing a negative 1D spring vector $x_{i+1,i}$).  

The $p=2$ case uniquely incorporates the factor-field into the original potential via the translations $x_{i+1,i} \mapsto x_{i+1,i}-b/k \, \forall i$:
\begin{align}
    \widetilde{U}(\widetilde{\bf x}, b) = \frac{k}{2}\sum_i \left(x_{i+1,i}-\frac{b}{k}\right)^2 \label{hc_factor2} + \textrm{const.} 
\end{align}
%Incorporating the factor-field via the translation in Eq. \eqref{hc_factor2} is unique to this model. 
This allowed previous work  %The factorisation scheme in Eq. \eqref{hc_factor2} is unique to the harmonic-chain model and allowed for previous work \cite{Lei2019ECMCwithFactorFields} 
to implement factor-field ECMC by sampling a single event per pairwise interaction for the $p=2$ case
~\cite{Lei2019ECMCwithFactorFields}. This is achieved by directly inverting Eq.~\eqref{ecmc_master_equation} for each pairwise interaction (with $U \mapsto k\left(x_{i+1,i}-b / k\right)^2/2$) and avoids the inflated mean event rate of the four-factor method. Cell-horizon ECMC provides the universal framework to generalise this benefit to all models.  In particular, it re-expresses the total potential of the generalised harmonic chain as
\begin{align}
    \widetilde{U}(\widetilde{\bf x}, b) &= \sum_i \left( \frac{k}{p} |x_{i+1,i}|^p - b x_{i+1,i} \right) , \label{hc_factor3}
\end{align}
which generalises the factorisation scheme defined by Eq.~\eqref{ff_event_rate} to all $p \ge 1$. Before demonstrating this approach for general $p \ge 1$, we explore the $p=2$ case as it provides analytic verification, including analytic predictions of $b^*$.

%In Appendix \ref{harm_chain_appendix_spring_length}, we show that the spring lengths are normally-distributed with $\mu_x = L/N$ and $\sigma^2_x=(\beta k)^{-1}(1-N^{-1})$. 
For $p=2$, both the mean event rate and the pressure can be used to predict $b^*$.  In Appendix \ref{harm_chain_appendix_event_rate}, we derive analytic expressions for the non-factorised mean event rate $\Lambda_{\rm nf} = \sqrt{\beta k / \pi} v_0$ and cell-horizon mean event rate 
\begin{equation}\label{event_rate_factor}
    \Lambda_{\rm ch} = \beta k v_0 \left[\Delta \text{erf}\left(\frac{\Delta}{\sqrt{2}\sigma_x}\right) + \sqrt{\frac{2}{\pi}}\sigma_x\exp\left(\frac{-\Delta^2}{2\sigma_x^2}\right)\right]
\end{equation}
where $\Delta = b/k - L/N$ and $\sigma_x^2$ is the variance of the spring lengths. $\Lambda_{\rm ch}$ is minimised at $b = Lk/N$ (i.e. $\Delta = 0$) and $\Lambda_{\rm ch}(b = Lk/N) > \Lambda_{\rm nf}$ for all $N > 2$, with $\lim_{N\rightarrow \infty}\Lambda_{\rm ch}(\Delta=0)/\Lambda_{\rm nf} = \sqrt{2}$. 
%Eqs~\eqref{event_rate_unfactor} and \eqref{event_rate_factor}, and the associated $b^*$ and large-$N$ limit, demonstrate excellent agreement with ECMC simulations (Fig. \ref{fig:hc_er_vs_b}). 
Excellent agreement between these predictions and ECMC simulations is demonstrated in Fig.~\ref{fig:HarmonicChain}(a) (which presents both as functions of $b$ and $N$ at fixed $\rho$; the inset presents $\Lambda_{\rm ch}(b = Lk/N) / \Lambda_{\rm nf}$ vs $N$ at fixed $\rho$ to emphasise the inequality).  In Appendix \ref{harm_chain_appendix_pressure}, we derive the analytic expression 
\begin{align}\label{eq:HarmChainPressure}
    \beta P = \beta \left( b - \frac{Lk}{N} \right) 
\end{align}
%where $v_{idx}$ is the velocity on the index space and $P(b^*=L/N) = v_{idx}(b^*=L/N) = 0$. 
%Numerical results in \blue{Fig \ref{fig:index_velocity_pressure}} demonstrate that 
for the normalised pressure.  Numerical results in Appendix~\ref{app:Vidx} are consistent with the index-space velocity $v_{\rm idx} = \beta \left(b - Lk / N\right) v_0$ ($v_{\rm idx}$ is the long-time displacement on the index space per unit time).  %Moreover, numerical results for other models (derived in \blue{Appendix XX}) are consistent with the ansatz $P \propto v_{\rm idx}$.  
Selecting $b = b^*_P$ (with $b^*_P$ defined such that $P(b=b^*_P) = 0$) therefore leads to %optimal dynamics on the index space, i.e. 
$v_{\rm idx} = 0$ or an equal number of forwards and backwards events on long time scales.  The hypothesis that $v_{\rm idx} = 0$ optimises autocorrelation then predicts that the autocorrelation time $\tau$ of the $q=1$ structure factor $S(q) = \left| \sum_j \exp(i q x_j) \right|^2 / N$ is minimised at $b=Lk/N$, as supported by the results in Fig. \ref{fig:HarmonicChain}(b) (which present $\tau$ as a function of $b$ and $N$ at fixed $\rho$).  Indeed, the analogous results for the computational effort presented in Fig. \ref{fig:HarmonicChain}(c) are minimised at $b = Lk/N$, reflecting the pressure and index-space velocity both correctly predicting $b^*$ for the $p=2$ case.  More generally, however, the two metrics do not predict the same $b^*$.  We discuss these subtleties in Section~\ref{sec:OptimalB} and measure sampling performance relative to $b=b^*_P$ throughout.

\begin{algorithm}[t]
\caption{Sampling event $k$ with cell-horizon ECMC for 1D models with positive singularities}\label{alg:cell_horizon_singularities}
\begin{algorithmic}
\Require %current positions vector $\mathbf{x} = \mathbf{x}(t_{k - 1})$, 
current springs vector $\widetilde{\mathbf{x}} = \widetilde{\mathbf{x}}(t_{k - 1})$, current active particle $a = a(t_{k - 1})$, %\blue{current active-particle position $x_a = x_a(t_{k - 1})$}, 
cell horizon $x_h$, factor-field $b$, inverse temperature $\beta$, number of particles $N$, speed $v_0$

\For{$j \gets 1$ \textbf{to} $N$}
    \If{$j = a-1$}
        \State $\widetilde\lambda^{\text{max}}_{aj} \leftarrow \widetilde\lambda_{aj}(x_{aj} + x_{a+1,a}, b)$ 
        \State $\eta^k_{aj} \sim \text{Exp}(\widetilde\lambda^{\text{max}}_{aj})$
    \Else
        \State Solve Eq. \eqref{ecmc_master_equation} for $\eta^k_{aj}$ with $U$ and $t_k$ replaced with $U_{aj}$ and $t^k_{aj}$, respectively. %$U$ replaced with $U_{aj}$. %$U=\widetilde{U}(x_{aj}, 0)$ given in Eq. \eqref{eq:soft_disk_potential}.
    \EndIf
\EndFor

\State $\eta_k \leftarrow \min(\{\eta^k_{aj}\})$ 
\State $w \leftarrow \text{arg}\min_j (\{\eta^k_{aj}\})$
%\State $x_a \leftarrow x_a + \eta_k$
\State $x_{aj} \leftarrow x_{aj} + \eta_k$ \textbf{for all} $j \ne a$ (i.e. $x_a \leftarrow x_a + \eta_k$)
\State $t_k \leftarrow t_{k - 1} + \eta_k / v_0$
\State $\Gamma \sim \mathcal{U}(0,1)$
    
\If{$w = a + 1$}
	\State $\widetilde{\lambda}_{aw}^{\rm max} \leftarrow \lambda_{aw}(x_{aw})$
\EndIf

\If{$w \ne a \pm 1$ \textbf{or} $\Gamma \leq \widetilde\lambda_{aw}(x_{aw}, b)/\widetilde\lambda^{\text{max}}_{aw}$}
    \State $a \leftarrow w$
\EndIf

\end{algorithmic}
\end{algorithm}

\begin{figure}[t]
    \includegraphics[width=\linewidth]{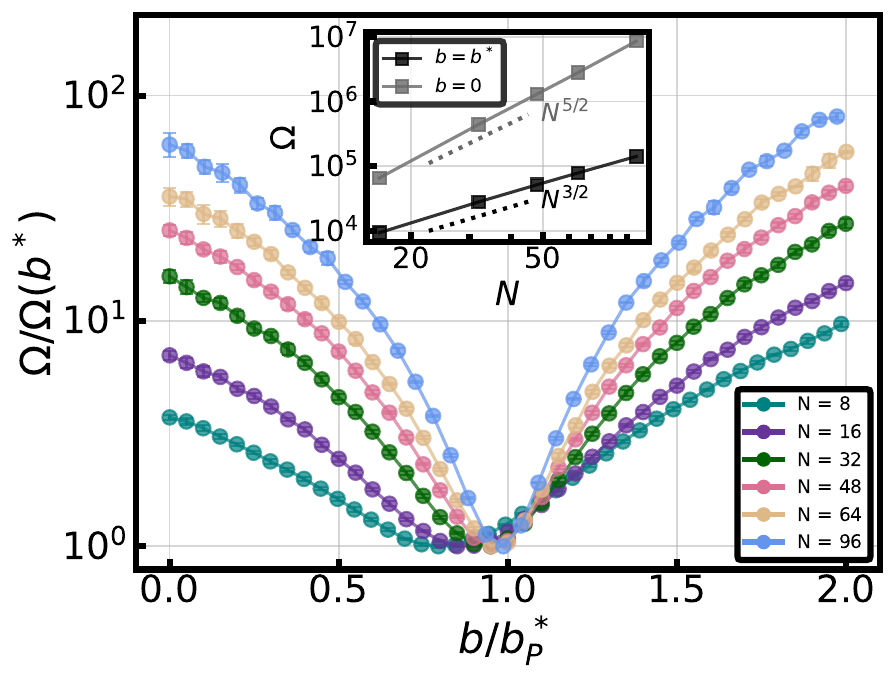}
    \caption{MC estimates of the computational effort $\Omega$ (defined in Section~\ref{sec:FactorFields}) of cell-horizon ECMC as a function of $b$ and $N$ for the $\kappa=2$ 1D soft-disk model with mean particle density $\rho = (3 / 4) / \sigma$.  Results are consistent with $\Omega$ being minimised at $b=b^*_P$ in the large-$N$ limit.  Inset: $\Omega$ as a function of $N$ at fixed $\rho = (3 / 4) / \sigma$ at $b = 0$ and $b = b^*$.  Results are consistent with $\Omega$ being $\mathcal{O}(N^{5/2})$ and $\mathcal{O}(N^{3/2})$ for the standard-factorised and optimal cell-horizon approaches.  This reflects cell-horizon ECMC hugely outperforming other methods by the key metric of sampling performance ($\Omega$), even in the presence of positive singularities and long-range interactions.  Lines are a guide to the eye. Results were averaged across 10-12 independent simulations, each of between $10^5$ and $5 \times 10^6$ samples (see Appendix~\ref{sec:CodeAndSimulationData} for precise details).}
 \label{fig:comp_effort_vs_b_soft_disks}
\end{figure}

\begin{figure*}[htb]
\includegraphics[width=\linewidth]{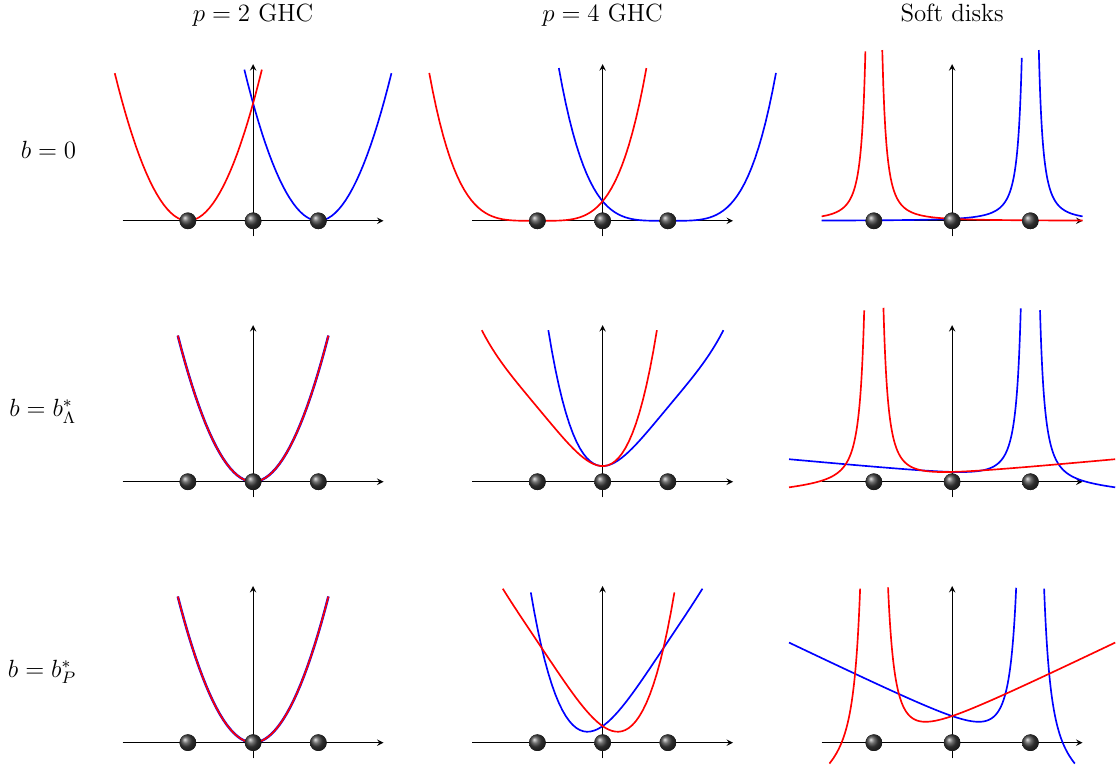}
\caption{Schematics illustrating the effect of key choices of $b$.  For modal state-space configurations, the functions $U_{a,a-1} - b x_{a,a-1}$ (red) and $U_{a+1,a} - b x_{a+1,a}$ (blue) are presented at $b=0$, $b=b^*_\Lambda$ and $b=b^*_P$ for both the generalised harmonic chain (GHC; $p=2$ and $p=4$) and 1D soft-disk model.  Green curves represent $U_{a,a-1} - b x_{a,a-1}$ and $U_{a+1,a} - b x_{a+1,a}$ functions that coincide and $y$-shifts have been applied for illustrative purposes.  For the $p=2$ GHC, $b^*_\Lambda = b^*_P$.  For all other models, the value of $b$ that minimises the mean event rate ($b^*_\Lambda$) does not coincide with that which induces optimal autocorrelation ($b^*_P$).}
\label{fig:potentials_at_different_b}
\end{figure*}

We now compare sampling performance between the non-factorised, standard-factorised (i.e. $b=0$) and cell-horizon strategies.  
%It is now informative to compare sampling performance between four key ECMC sampling strategies: non-factorised, standard-factorised (i.e. $b=0$), four-factor and cell horizon.  
Fig \ref{fig:HarmonicChain}(c) presents the computational effort as a function of $b$ and $N$ for the non-factorised and cell-horizon strategies at fixed mean particle density $\rho$ (the cell-horizon results encompass standard-factorised ECMC at $b=0$).  Whilst standard-factorised ECMC exhibits a larger computational effort than the non-factorised process at the presented values of $N$, cell-horizon ECMC is more efficient when $b$ is chosen optimally.  In addition, Fig. \ref{fig:HarmonicChain}(d) presents the autocorrelation time and computational effort of the non-factorised, standard-factorised and optimal cell-horizon (i.e. $b = b^*$) approaches as a function of $N$ at fixed $\rho$. The results are consistent with the methods scaling as $\mathcal{O}(N^3)$, $\mathcal{O}(N^{5/2})$ and $\mathcal{O}(N^{3/2})$, respectively.  This additionally demonstrates that even standard-factorised ECMC is more efficient than the non-factorised approach in the large-$N$ limit, due to the latter inducing non-local jumps of $a$ on $\mathcal{I}$.  This is an example of factorisation not just being a tool for tractability, but also leading to higher sampling efficiency at large $N$ -- and for all $N$ in the case of optimal cell-horizon ECMC.

The key contributions of the cell-horizon strategy are i) to circumvent the impractically large computational effort of four-factor ECMC at low/high mean particle density for attractive/repulsive interactions (i.e. where $b^*$ is large), and ii) to provide a general analytic method for the factorisation scheme defined by Eq.~\eqref{ff_event_rate}.  Both are consistent with the results in Fig.~\ref{fig:harm_chain_figures_2}.  Fig. \ref{fig:harm_chain_figures_2}(a) presents the computational effort of the four-factor and cell-horizon approaches at $b = b^*$ as a function of mean particle density $\rho$ (at fixed $N$) for both the $p = 2$ and $p = 4$ cases.  For $p=2$, the four-factor results scale as $\mathcal{O}(\rho^{-2})$ whereas the cell-horizon results are insensitive to particle density.  For $p=4$, the results become $\mathcal{O}(\rho^{-4})$ and $\mathcal{O}(\rho^{-1})$, respectively.  This reflects cell-horizon ECMC circumventing the impractically large computational effort of four-factor ECMC at large $b^*$ (at small $\rho$ in this case), curtailing its linear increase in mean event rate (with $b$).  The relative efficiency with respect to decreasing particle density increases with the power $p$.  In addition, Fig.~\ref{fig:harm_chain_figures_2}(b) presents the $p=4$ autocorrelation time and computational effort as a function of $N$ and at fixed $\rho$ for the four-factor and cell-horizon strategies.  Both sets of results are $\mathcal{O}(N^{3/2})$, with a significantly smaller prefactor on the cell-horizon results.  Taken together, this is an example of cell-horizon ECMC providing a universal analytic framework for simulating the factorisation scheme defined by Eq.~\eqref{ff_event_rate}.  %\blue{More precisely, the original exposition of factor-field ECMC exploited Eq.~\eqref{hc_factor2} to achieve similar computational effort for the $p=2$ harmonic chain, but cell-horizon ECMC generalises to all translation-invariant pairwise models on the torus.}

\subsection{Soft disks}
\label{sec:SoftDisks}

The total potential of the soft-disk model on the 1D torus of length $L$ is defined as
\begin{equation}\label{eq:soft_disk_potential}
    \widetilde{U}(\widetilde{\bf x}, b) = \epsilon \sum_{j,i>j}\left(\frac{\sigma}{\mathfrak{d}(x_{i,j})}\right)^\kappa - b \sum_i x_{i+1,i} ,
\end{equation}
where $\epsilon > 0$ is an energy scale, $\sigma > 0$ is the diameter of the disk in the large limit $\kappa$ (hard-disk), $\sum_i x_{i+1,i}=L$ and $\mathfrak{d}(x_{ij}) := |(x_{ij} + L/2) \mod{L} - L/2|$ is the minimal separation distance between the particles $i$ and $j$.  As the particles are indistinguishable and $\widetilde{U}(\widetilde{\bf x}, b=0)$ is a function only of the minimal separation distances, we apply the additional constraint $x_{i + 1, i} > 0$ for all $i$ to avoid a countably infinite number of configurations with equal $\widetilde{U}(\widetilde{\bf x}, b=0)$.

The soft-disk potential features singularities at $x_{ij}=0$. This makes the bounding of the event rate required for Alg. \ref{alg:cell_horizon} impossible, and the approach must be modified. Alg. \ref{alg:cell_horizon_singularities} 
outlines the modified procedure. 
The singularities additionally fix the particle ordering on the state space under event-chain dynamics.  The position of the forwards neighbour therefore defines the position of the cell horizon of the backwards neighbour.  For all the other particles $j$, we sample events by replacing %$U$ in Eq.~\eqref{ecmc_master_equation} with the associated pairwise potential $U_{aj}$. 
(respectively) $U$ and $t_k$ with $U_{aj}$ and $t^k_{aj}$ in Eq.~\eqref{ecmc_master_equation}.  For the forwards neighbour, this has the effect of bounding the event rate $\beta v_0 \max\left(0, \partial_a U_{a+1,a} + b\right)$ (see Eq.~\eqref{ff_event_rate} and note that $b^* < 0$ for this repulsive model) with the $b=0$ event rate $\beta v_0 \max\left(0, \partial_a U_{a+1,a} \right)$. The prescription outlined in Section~\ref{sec:CellHorizonFramework} is then followed to define the next event distance $\eta_k$ and vetoing particle $w$.  If $w$ is the forwards or backwards neighbour, it is similarly confirmed as the new active particle by thinning. %the thinning process outlined in Section~\ref{sec:CellHorizonFramework} is followed to decide whether $w$ becomes active.  
This generalises cell-horizon ECMC to models containing repulsive interactions, positive singularities and long-range interactions.  %Neighbouring particles are partitioned into backwards and forwards neighbours based on their initial positions relative to the active particle. For backwards neighbours, we bound the event rate by considering the furthest distance that the active particle could travel. It is impossible for the active particle to move past the position of its closest forwards neighbour due to the singularity in the potential. The distance from the active particle to this neighbour is identified and used to compute the bounding rate. For forward neighbours, event distances are drawn by directly inverting Eq. \ref{ecmc_master_equation} with $U=U_{sd}(x_{a,j}, 0)$. Once events have been sampled for each neighbour, the minimum distance is taken as the final event distance and the associated particle taken as the vetoing particle. An accept-reject step is then performed to decide whether the active or vetoing particle is active for the next move. With this approach, we generalise the reduction in the computational effort previously presented for attractive potentials.% 

\begin{figure*}
    \centering
    \includegraphics[width=\linewidth]{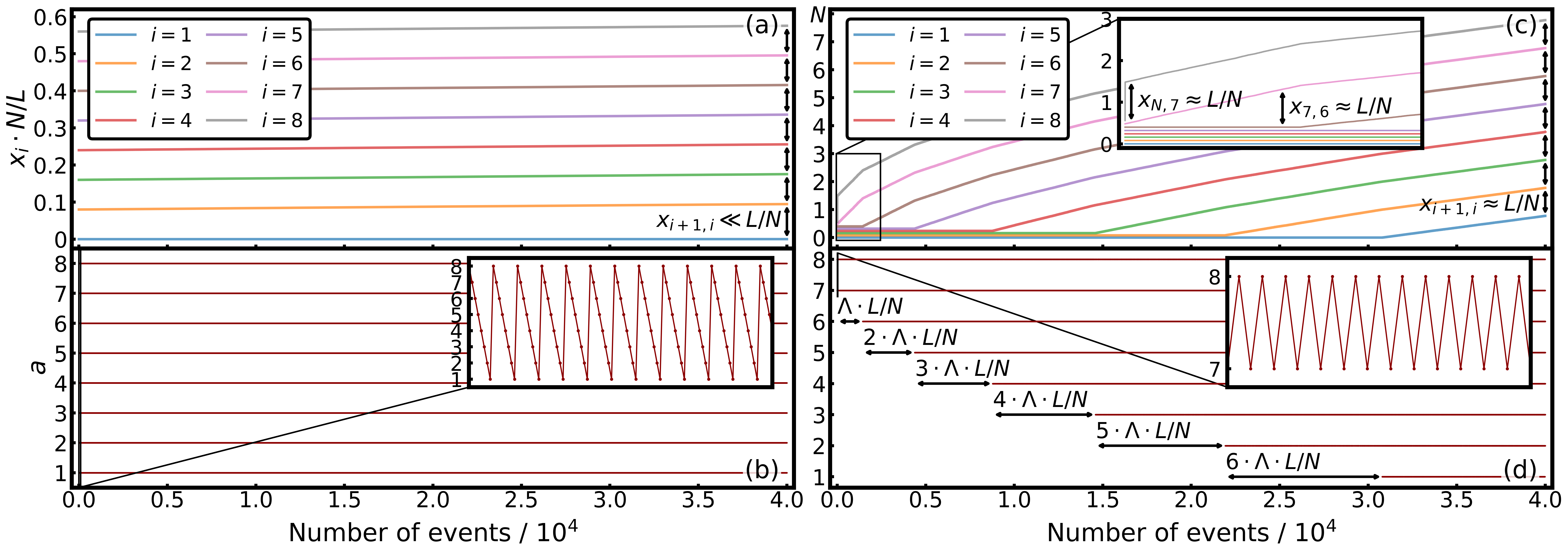}
    \caption{Trace plots of the particle positions $\{x_i\}$ (a,c) and active-particle index $a$ (b,d) of the $p=2$ generalised harmonic chain at strong coupling ($\beta k / \rho^2 = 4 \times 10^6$) and for $N = 8$ particles -- using standard-factorised ECMC, i.e. $b=0$ (a,b) and optimal cell-horizon ECMC, i.e. $b=b^*$ (c,d).  Simulations start from a single high-density cluster of particles (i.e. from a low-probability region corresponding to $\mathfrak{d}(x_{ij}) \ll L / N$ for all pairs $(i,j)$) with $a$ set to $N$ (i.e. at the edge of the cluster).  (a,b) For $b=0$, the backwards neighbour becomes active following each event, resulting in $a$ drifting deterministically through $\mathcal{I}$.  After $4\times10^4$ events, it still holds that $\mathfrak{d}(x_{ij}) \ll L / N$ for all pairs $(i,j)$.  (c,d) For $b = b^*$, particle $N$ (which is initially active) advances on the state space until it reaches a distance $L/N$ from its backwards neighbour $(i=N-1)$.  $a$ then alternates between these two activated particles until particle $N-1$ reaches a distance $L/N$ (on the state space) from its backwards neighbour $(i=N-2)$ which subsequently becomes active.  $a$ then cycles back and forth through the set of three activated particles until particle $N-3$ becomes activated.  This process repeats until $x_{i+1,i} \approx L / N$ for all $i$ (achieved after $4\times10^4$ events).}
    \label{fig:mixing_traj_bstar}
\end{figure*}

In Appendix~\ref{harm_chain_appendix_pressure}, we present the normalised pressure 
\begin{equation}\label{eq:SoftDiskPressure}
    \beta P = \frac{N}{L} \left(1 + \frac{\beta\kappa\langle \widetilde{U}(b=0) \rangle - 1}{N} \right) + \beta b ,
\end{equation}
showing that $b_P^*=(1-\beta\kappa\langle \widetilde{U}(b=0) \rangle - N
)/(\beta L)$. %($\langle \cdot \rangle$ denotes a thermal average, i.e. the expectation with respect to the target distribution).   
Fig.~\ref{fig:comp_effort_vs_b_soft_disks} presents the computational effort of cell-horizon ECMC applied to the $\kappa = 2$ soft-disk model as a function of $b$ and $N$ at fixed $\rho$.  The inset demonstrates that the method successfully achieves the $\mathcal{O}(N^{3/2})$ scaling at $b=b^*$, in contrast with the $\mathcal{O}(N^{5/2})$ scaling of the $b=0$ case. %, where these scaling results assume the cell-veto method~\cite{Kapfer2016Cell-vetoMCAlgorithm}.
The results presented in the main figure are consistent with $b = b^*_P$ minimising the computational effort in the large-$N$ limit.  The results at small $N$, however, demonstrate finite-size corrections to this prediction.  %Moreover, we found that the autocorrelation time is minimised by the same $b$ values.  Taken together, this \blue{is consistent with the hypothesis that} i) the computational effort is dominated by the autocorrelation time \blue{(as also supported by the results in Figs~\ref{fig:HarmonicChain}(b-c)) }, and ii) zero drift on the index space (as ensured by $b = b^*_P$; \blue{see Fig. XX}) can only guarantee optimal autocorrelation at large $N$.  
%Zero drift on the index space (as ensured by $b = b^*_P$; \blue{see Fig. XX}) 
Choosing $b = b^*_P$ can therefore only guarantee optimal autocorrelation in the large-$N$ limit.  

%\blue{Selecting $b=b^*_\tau$ nullifies the pressure and cancels the drift on the index space. The factor-field $b^*$ that minimises the computational effort was found to approach $b^*_\tau$ at large $N$. For smaller systems, however, the effect of the choice of $b$ on the mean event rate was significant, and it was instead optimal to choose $b<b^*_\tau$, i.e. closer to $b^*_\Lambda$ to reduce the mean event rate (Fig. \ref{fig:comp_effort_vs_b_soft_disks}).  This is due to the mean event rate being relatively large at $b^*_\tau$ (compared with its value at $b^*_\Lambda$) but $\tau$ dominating at large $N$ ($\tau$ is $\mathcal{O}(N^{3/2})$ but the mean event rate is only weakly dependent on $N$).  These results emphasise that a compromise between minimising the mean event rate and the pressure is required for optimal sampling.}

\subsection{Optimal $b$}
\label{sec:OptimalB}

%Moreover, numerical results for other models (derived in \blue{Appendix XX}) are consistent with the ansatz $P \propto v_{\rm idx}$.

The $p=2$ harmonic chain exhibits a special case where a single value of $b$ optimises both the mean event rate and autocorrelation.  The mean event rate is governed by the dynamics of the active particle on the state space \textit{between events}, whereas the autocorrelation is a product of the dynamics on the index space \textit{at events}.  In general, the value of $b$ that minimises the mean event rate does not necessarily induce zero drift on the index space (our numerical results support the latter being optimal for the autocorrelation at large $N$).  This leads to an additional value $b^*_\Lambda$ that minimises the mean event rate and is distinct from $b^*_P$ (numerical results in Appendix~\ref{app:Vidx} are consistent with the general ansatz $\beta P = v_{\rm idx} / v_0$, i.e. $v_{\rm idx} = 0$ at $b = b^*_P$ in general).
 
The generalised harmonic chain with $p\neq2$ provides an example of $b^*_\Lambda \ne b^*_P$.  Choosing $b$ to translate the minima of $U_{a+1,a} - b x_{a+1,a}$ and $U_{a,a-1} - b x_{a,a-1}$ to the modal position of the active particle relative to $x_{a \pm 1}$ (which minimises the mean event rate) results in different gradient profiles in each case (see Fig. \ref{fig:potentials_at_different_b}). There will consequently be an index-space drift accompanied by a non-optimal $\tau$. Ultimately, the computational effort must be relied on as the general metric of sampling efficiency. For the $p=4$ generalised harmonic chain, we have found that the computational effort and autocorrelation time are both minimised by choosing $b=b^*_P$ to induce zero index-space drift.  Similarly, the computational effort and autocorrelation of the soft-disk model are minimised by the same $b$ value for any given model parameters (though this coincides with $b^*_P$ only in the large-$N$ limit).  The computational effort is therefore dominated by the autocorrelation time and $b^*_P$ is the large-$N$ predictor of $b^*$.  %This is explained by the scale of the improvement in the autocorrelation time typically being larger than the improvement in the mean event rate.

\section{Mixing}
\label{sec:Mixing}

The capacity of factor-fields to optimise autocorrelation was ascribed to superior lifting dynamics that promptly relax local density variations. To illustrate this, Fig.~\ref{fig:mixing_traj_bstar} presents trace plots of both the particle positions and active-particle index $a$ of the $p=2$ generalised harmonic chain at strong coupling ($\beta k / \rho^2 = 4 \times 10^6$) and for $N = 8$ particles.  We initialise from a single high-density cluster of particles (i.e. from a low-probability region) and simulate using both standard-factorised ECMC (i.e. $b=0$) and optimal factor-field ECMC (i.e. $b=b^*_P$) and with $a$ initially set to $N$ (i.e. at the edge of the cluster).  %This is the most extreme example of the density variations that are efficiently relaxed at equilibrium with $b = b^*$.

For $b=0$, the backwards neighbour becomes active following each event. This results in $a$ drifting deterministically through $\mathcal{I}$ and leads to sub-optimal mixing on the original state space (Figs~\ref{fig:mixing_traj_bstar}(a,b)). In contrast, utilising the optimal factor-field $b^*_P=Lk/N$ (Figs~\ref{fig:mixing_traj_bstar}(c,d)) results in flexible index-space dynamics that lead to the most efficient dynamics on the original state space. The particle at the edge of the cluster $(i=N)$ advances on the state space until it reaches a distance $L/N$ from its backwards neighbour $(i=N-1)$. %The active-particle index 
$a$ then alternates between these two activated particles until particle $N-1$ reaches a distance $L/N$ from its backwards neighbour $(i=N-2)$ which subsequently becomes active. %\blue{In the strong-coupling limit, the active particle index now advances forwards through this set of three activated particles until it reaches $N$ then reverses and advances backwards until it reaches $N-2$ and reverses again to repeat the process, replicating the superdiffusive excursions described at equilibrium in Section~\ref{sec:FactorFields}. This continues until particle $N-3$ becomes activated (when $x_{N-2,N-3}$ reaches $L/N$) and the entire process repeats until all particles have been activated.} 
%The active-particle index 
$a$ then cycles back and forth through the set of three activated particles until particle $N-3$ becomes activated.  This repeats each time the number of activated particles $N_{\rm activated}$ increases.  Once $N = N_{\rm activated}$, %the minimal state-space separation distances between neighbouring particles are each approximately $L/N$ 
$x_{i+1,i} \approx L / N$ for all $i$ and the local density variation has been relaxed, i.e. the efficient collective particle dynamics have relaxed the system. We tend to observe extended periods of $a$ advancing through the set of activated particles in a single direction, %providing exaggerated examples of 
replicating the superdiffusive excursions described at equilibrium in Section~\ref{sec:FactorFields}.  This mixing scenario elucidates the efficient relaxation of less extreme density variations that arise frequently at equilibrium.

The results in Fig.~\ref{fig:mixing_traj_bstar}(d) demonstrate that the number of events required to activate the next non-activated particle scales with $N_{\rm activated}$, leading to an $\mathcal{O}(N^2)$ mixing time in terms of events.  By contrast, $N_{\rm activated}$ reaches $N$ after $\mathcal{O}(N)$ events in the $b=0$ case (see Fig.~\ref{fig:mixing_traj_bstar}(b)) but leaves the particles in a similar configuration on the original state space (see Fig.~\ref{fig:mixing_traj_bstar}(a)).  
%This a significant improvement on the $\mathcal{O}(N^{5/2})$ mixing time that we measured for the $b=0$ case, which has a very large prefactor due to the high number of events per unit distance travelled evidenced in Fig. \ref{fig:mixing_traj_b0}.
This results in an $\mathcal{O}(N^{5/2})$ mixing time (see Appendix~\ref{app:Vidx}) with a very large prefactor due to the high number of events per unit distance travelled, providing further evidence that less efficient exploration of the index space can lead to much more efficient exploration of the original state space.

\section{Discussion}
\label{sec:Discussion}

%Existing non-numerical strategies for generalising factor-field ECMC require the four-factor approach~\cite{Krauth2024HMCvsECMC}. 
Previous non-numerical strategies for generalising factor-field ECMC have focussed on the four-factor approach~\cite{Krauth2024HMCvsECMC}.  This achieved the groundbreaking $\mathcal{O}(N^{3/4})$ improvement on the computational effort of MD and HMC for translationally symmetric pairwise models (of nearest-neighbour interactions) on the 1D torus, but uses a factorisation scheme with mean event rate that is linear in the factor-field $b$. This results in an impractically large prefactor on the computational effort at low/high mean particle density for attractive/repulsive interactions, due to a high optimal $b$. Our approach, by contrast, provides an analytic framework for sampling events with respect to the optimal factorisation scheme defined by $\widetilde{\lambda}_{ij}$ (see Eq.~\eqref{ff_event_rate}). This circumvents the linear relationship between the mean event rate and $b$, resulting in a universal sampling strategy with computational effort that is robust to mean particle density $\rho$. % and model.  
For example, our method is insensitive to $\rho$ for the $p = 2$ generalised harmonic chain -- an attractive model -- as opposed to the $\mathcal{O}(\rho^{-2})$ scaling of the four-factor strategy. This improvement increases for the more challenging $p = 4$ case, with respective $\mathcal{O}(\rho^{-1})$ and $\mathcal{O}(\rho^{-4})$ scalings.  %We additionally elucidated the efficiency of factor-field ECMC by demonstrating how its efficient collective particle dynamics relax local density variations on fast timescales (via simulations initialised far from equilibrium at $b=b_\tau^*$).  We combined this with the surprising result that even standard two-particle factorisation produces an $\mathcal{O}(N^{1/2})$ improvement on the computational effort with respect to non-factorised ECMC -- as the latter cannot induce the efficient collective particle dynamics.  
We additionally presented the surprising result that even standard two-particle factorisation produces an $\mathcal{O}(N^{1/2})$ improvement on the computational effort of non-factorised ECMC -- as the latter induces non-local jumps of the active-particle index $a$ on the index space $\mathcal{I}$, hindering its ability to relax local density variations via collective particle motion.  By contrast, optimal factor-field ECMC induces an efficient collective particle dynamics that relax local density variations on fast timescales, which we demonstrated via simulations initialised far from equilibrium.  The factorisation scheme defined by $\widetilde{\lambda}_{ij}$ is therefore the optimal ECMC sampling strategy for translationally symmetric pairwise models on the torus, with the cell-horizon scheme providing a universal framework to achieve this in general (the concept of \emph{particle lanes} extends to the higher dimensional torus~\cite{Maggs2022LargeScaleDynamicsOfECMC}).  We also introduced computational effort as a metric for sampling performance and elucidated the index-space velocity as a predictor for optimising this at large $N$, numerically demonstrating its agreement with predictions made by the pressure.

The concept of index-neighbour models provides the natural route to expanding this framework to lattice spin models such as spin glasses and the 2DXY model.  For the latter, it will be interesting to extend previous studies on the dynamics of the active-particle location~\cite{Kimura2017AnomalousDiffusion} and develop theory on behaviour at criticality, with broad potential impact from continuous phase transitions to %Wick-rotated approaches to quantum field theory~\cite{Edwards2019Applications,Bazavov2010NonperturbativeQCD}.  
worldline quantum field theory~\cite{Edwards2019Applications} and lattice QCD~\cite{Bazavov2010NonperturbativeQCD}.
Moreover, extensions to multiple species will provide a route to exploiting cell-horizon ECMC in the context of particle models of glasses, where the combination with teleportation portals~\cite{Bierkens2023PDMPsWithBoundaries} could prove powerful~\cite{Lenton2026GlassyECMC}.  In the longer term, extensions to molecular rotations and many-body interactions~\cite{Harland2017EventChain} may accelerate simulations of complex molecular models, from atomistic water to peptides in solution.  Incorporating enhanced sampling functionality (e.g. tempering) could then address challenging multimodal problems such as peptide folding and the sampling of %distinct crystalline phases formed by liquid crystallisation.
competing crystalline phases formed by off-lattice particle models.  It will also be natural to exploit cell-horizon ECMC to extend recent work on the dynamics of standard-factorised ECMC across phase-separated boundaries~\cite{Tartero2026LiftingTheFog}, building further impact at phase transitions.

Outside the physical sciences, similar factorisation schemes have been combined with control variates and subsampling to leverage significant performance improvements when applying the zigzag process to large data sets in Bayesian inference~\cite{Bierkens2019TheZigZagProcess}.  Exploiting possible connections between control variates and factor-fields may provide advances in both fields, %e.g. Bayesian analogues of the pressure characterise model misspecification.  
e.g. nullifying the pressure is analogous to minimising Bayesian model misspecification~\cite{Faulkner2024SamplingAlgorithms}.  In addition, future work generalising ECMC with broken velocity symmetry to non-toroidal compact state spaces may provide the framework for broadening cell-horizon ECMC to the Strauss process, stochastic SIR model and other models in computational statistics.  Finally, these results are an example of less efficient exploration of the auxiliary (index) space leading to much faster mixing on the original state space.  It may be illuminating to characterise this in terms of large-deviations theory. 

\begin{acknowledgments}

We are grateful to W. Krauth and S. Livingstone for helpful discussions.  J.~G. is supported by an EPSRC studentship funded by the Centre for Doctoral Training in Modelling of Heterogeneous Systems EP/S022848/1.  M.~F.~F. is supported by EPSRC grant UKRI4536.  Simulations were performed using the Scientific Computing Research Technology Platform at the University of Warwick.

\end{acknowledgments}

\appendix 

\onecolumngrid

\section{Code and simulation data}
\label{sec:CodeAndSimulationData}

Code will be made freely available on GitHub at \href{https://github.com/michaelfaulkner/super-aLby}{https://github.com/michaelfaulkner/super-aLby} following the review process (the README will outline how to reproduce all published data).  All simulation data will be made available at a data repository.  Simulations for Figs~\ref{fig:HarmonicChain}--\ref{fig:comp_effort_vs_b_soft_disks} and \ref{fig:index_velocity_pressure} started from ordered configurations.  Those for Figs~\ref{fig:mixing_traj_bstar} and \ref{fig:MixingTimes} started far from equilibrium.  The trace plots in Fig.~\ref{fig:mixing_traj_bstar} were reconstructed from a skeleton formed at events.  All samples of the structure factor were formed by recording its configurational value with some frequency $\propto 1/N$.  We then corrected for the $1/N$ factor after calculating the estimated autocorrelation time $\tau$ of each sample (and similarly for the mixing times presented in Appendix~\ref{app:Vidx}).  This circumvents memory constraints when simulating large numbers of particles.  Results were then averaged across 10-12 independent simulations, each of between $10^5$ and $5 \times 10^6$ samples (the above README will provide the exact simulation details of each figure).  $\tau$ errors were estimated using standard error estimates across the independent simulations, with all non-visible error bars smaller than the marker size.

\section{Breaking velocity symmetry}\label{breaking_vel_symm}

For models on the 1D torus $\mathbb{T}$, ECMC augments the original state space %$\mathbb{T}^N$ 
$\mathbb{R}^{N(N - 1) / 2}$ (of the springs vector $\widetilde{\bf x}$) with both the \emph{velocity space} $\mathcal{V} = \{ -v_0, v_0\}$ and index space $\mathcal{I}$.  This results in the joint distribution %$\pi(\mathbf{x}) \mu(v) \nu(a)$ 
$\pi(\widetilde{\bf x}) \mu(v) \nu(a)$ over the augmented state space %$\mathbb{T}^N \times \mathcal{V} \times \mathcal{I}$, 
$\mathbb{R}^{N(N - 1) / 2} \times \mathcal{V} \times \mathcal{I}$, where the \emph{velocity distribution} $\mu(v)$ is uniformly distributed on $\mathcal{V}$ and the \emph{index distribution} $\nu(a)$ is uniformly distributed on $\mathcal{I}$.  Breaking velocity symmetry is then defined as constructing a continuous-time Markov process that samples only half of $\mathcal{V}$, i.e. one that fixes $v$ to either $+v_0$ or $-v_0$ throughout the process on the 1D torus (n.b. the definition also applies to the higher dimensional torus).  

In the case of ECMC without velocity changes, the state of the Markov process reduces to %$(\mathbf{x}, a)\in \mathbb{T}^N \times \{1, \ldots ,N\} $,
%\blue{$(\mathbf{x}, a)\in \mathbb{T}^N \times \mathcal{I}$}, with invariant probability measure written with a slight abuse of notation as $\pi ( \mathbf{x}, a) := \pi(\mathbf{x})/N$. 
$(\widetilde{\bf x}, a)\in \mathbb{R}^{N(N - 1) / 2} \times \mathcal{I}$, with invariant probability measure written with a slight abuse of notation as $\pi ( \widetilde{\bf x}, a) := \pi(\widetilde{\bf x})/N$.  For a translationally symmetric model of particles on $\mathbb{T}$ (with springs vector on $\mathbb{R}^{N(N - 1) / 2}$), here we present an informal argument that 
\begin{align}\label{eq:StationarityAim}
    \pi P^{h} (\widetilde{\bf x}, a)=\pi(\widetilde{\bf x}, a) + \mathcal{O}(h^2)\ 
\end{align}
for small $h > 0$, i.e. that broken velocity symmetry preserves ergodicity on the original state space, and where $P$ denotes the transition semigroup of the symmetry-broken algorithm that sets $v(t) = v_0$ for all $t$.  We shall present the proof for non-factorised ECMC though an almost identical argument can be used for the factorised approaches.

\def\dUa{\partial_a U}
\def\dUap{\partial_{a'} U}
\def\dUapp{\partial_{a''} U}
\def\bx{\widetilde{\bf x}}
We denote by $e_{a}$ the unit vector in the $a^{\rm th}$ particle direction such that $\widetilde{\bf x} + h v_0 e_a$ adds $h v_0$ to $x_{a,a-1}$ and subtracts $h v_0$ from $x_{a+1,a}$ (recall that $\widetilde{\bf x} := (x_{1,2}, \dots , x_{1,N}, x_{2,3}, \dots , x_{2,N}, \dots, x_{N - 1, N})^T$).  Using $\exp \left( - \beta v_0 \max \{\dUa(\bx ),0\} h \right) = 1 - h \beta v_0 \max \{\dUa(\bx ),0\} + \mathcal{O}(h^2)$, the left-hand side of Eq.~\eqref{eq:StationarityAim} can be written as 
$$
\pi P^{h} (\widetilde{\bf x}, a) =
\pi (\widetilde{\bf x} - h v_0 e_a, a)(1 - h 
\beta v_0 \max \{\dUa(\bx ),0\})
+\sum_{a'} \pi(\bx, a')
h \beta v_0 \max \{ \dUap(\bx )  ,0\} 
 {
%\pi(\bx, a')
\max \{-\dUa(\bx ),0\}
\over
\sum_{a''}\max \{-\dUapp(\bx ),0\}
}
+\mathcal{O}(h^2)
$$
Now noting that 
$$\pi(\bx-hv_0 e_a, a) = \pi (\bx ,a) (1+h \beta v_0 \dUa(\bx ))+\mathcal{O}(h^2)\ 
$$
and using $\pi (\bx , a) = \pi (\bx ) / N$ for all $a$, we find that
\begin{align}
    \pi P^{h} (\widetilde{\bf x}, a) = {1\over N} \pi (\bx )
        \left[1 + h \beta v_0 \left( \dUa(\bx ) - \max \{\dUa(\bx ),0\} + \sum_{a'} \max \{ \dUap(\bx )  ,0\} 
        {\max \{-\dUa(\bx ),0\} \over \sum_{a''}\max \{-\dUapp(\bx ),0\}} \right) \right] + \mathcal{O}(h^2) . \nonumber
\end{align}
Ignoring terms of $\mathcal{O}(h^2)$, stationarity then requires that
%Ignoring terms of $O(h^2)$, and equating this expression to $\pi (\bx, a)$, so that stationarity requires that 
\begin{equation}
\label{e:stat1} 
 \dUa(\bx ) - \max \{\dUa(\bx ),0\} + \sum_{a'} \max \{ \dUap(\bx )  ,0\} 
 {
\max \{-\dUa(\bx ),0\}
\over
\sum_{a''}\max \{-\dUapp(\bx ),0\}
}
=0 ,
\end{equation}
where we have used $\pi P^{h} (\widetilde{\bf x}, a) = \pi (\bx, a)$ at stationarity and $\pi (\bx, a) = \pi (\bx) / N$ for all $a$.  Now we use the fact that $U(\bx)=\sum_{j,i>j} U_{ij}(x_{ij}) $ which implies that
$$
\sum_j \partial_j U(\bx) = 0 .
$$ %and also
Then also using
$$
\max \{\dUapp(\bx ),0\} - \max \{-\dUapp(\bx ),0\}
= \dUapp(\bx ) ,$$
%Therefore
we find that
$$
\sum_{a''}\max \{\dUapp(\bx ),0\} = \sum_{a''}
(\dUapp(\bx ) +\max \{-\dUapp(\bx ),0\})
= 0 +\sum_{a''}\max \{-\dUapp(\bx ),0\} .
$$
Thus
$$
\sum_{a''} \max \{\dUapp(\bx ),0\} = \sum_{a''}\max \{-\dUapp(\bx ),0\}
$$
and therefore the left-hand side of Eq.~(\ref{e:stat1}) reduces to 
$$
\dUa(\bx ) - \max \{\dUa(\bx ),0\} + \max \{-\dUa(\bx ),0\} = 0 ,
$$
as required.

\begin{comment}
We now note that since $\pi $ is bounded and %the state space is compact
\red{the springs vector $\bx$ is constrained by $\sum_i x_{i + 1, i} = L$}, $\pi (\bx , a)$ is a finite measure, so can be normalised to be a probability measure.

\red{**the red text above used to state ``the state space is compact'' which is correct for models defined in terms of particle positions on the torus, but we have to define in terms of 1D spring vectors each on $\mathbb{R}$ - replacement ok?**}

%In this appendix, we use $\langle \cdot \rangle$ to denote a thermal average, i.e. the expectation with respect to the target distribution.
\end{comment}

\section{Generalised harmonic chain}\label{harm_chain_appendix}

\subsection{Spring-vector distribution ($p=2$ case)}\label{app:SpringVecDistr}

Let $\{y_{i+1,i}\} \in \mathbb{R}$ be a set of iid unconstrained 1D spring vectors (each with distribution $\propto \exp(-\beta k y_{i+1,i}^2 / 2)$) and write $\sum_iy_{i+1,i}=S$, where $S$ is a random variable with $\langle S \rangle = 0$. Clearly, $\langle y_{i+1,i} \rangle = 0$ and the variance $\sigma^2_y = (\beta k)^{-1}$ by the equipartition theorem.  We then express the constrained spring vectors $\{x_{i+1, i}\}$ (of the $p = 2$ generalised harmonic chain on the 1D torus of length $L$) as:
\begin{equation}
    x_{i+1,i} = y_{i+1,i} - \frac{1}{N}\sum_j y_{j+1,j} + \frac{L}{N} %= y_{i+1,i} - \frac{1}{N}S + \frac{L}{N}
\end{equation}
This enforces the condition $\sum_i x_{i+1,i} = L$ but with $\langle x_{i+1,i} \rangle = L/N$.  Writing the right-hand side as 
\begin{align}
    %y_{i+1,i} - \frac{1}{N}S + \frac{L}{N} = 
    y_{i+1,i} - \frac{1}{N}\sum_j y_{j+1,j} + \frac{L}{N} = \left( 1 - \frac{1}{N} \right) y_{i+1,i} - \frac{1}{N}\sum_{j\ne i} y_{j+1,j} + \frac{L}{N}
\end{align}
allows us to determine the variance 
\begin{equation}
    \sigma_x^2 = \text{Var}(x_{i+1,i}) = \left( 1 - \frac{1}{N} \right)^2 \text{Var}(y_{i+1,i}) + \frac{1}{N^2}\text{Var}\left(\sum_{j\ne i} y_{j+1,j}\right) 
\end{equation}
since $\text{Cov}(y_{i+1,i}, y_{j+1,j}) = 0$ for all $i\ne j$.  It then follows that
\begin{equation}\label{eq:HarmChainVariance}
    \sigma_x^2 = \left( 1 - \frac{1}{N} \right)^2 \frac{1}{\beta k} + \frac{1}{N^2}(N-1)\frac{1}{\beta k} = \frac{1}{\beta k}\left(1-\frac{1}{N}\right) .
\end{equation}
The constrained spring vectors are therefore distributed as $x_{i+1, i}\sim \mathcal{N}(L/N, (\beta k)^{-1}(1-N^{-1}))$ for all $i$.

\subsection{Mean event rate ($p=2$ case)}\label{harm_chain_appendix_event_rate}
\subsubsection{Non-factorised ECMC}

The gradient of the total potential $\widetilde{U}$ (the $p=2$ case of Eq.\eqref{hc_factor3}) with respect to the position $x_i$ of particle $i$ is
\begin{equation}
    \frac{\partial \widetilde{U}}{\partial x_i} = k(x_{i,i-1}-x_{i+1, i}) .
\end{equation}
Writing $z_i := k(x_{i,i-1}-x_{i+1, i})$ defines a random variable with mean $\langle z_i \rangle = 0$ and variance 
\begin{equation}
    \sigma_z^2 = \text{Var}(z_i) = k^2[\text{Var}(x_{i, i-1}) + \text{Var}(x_{i+1, i}) - 2 \text{Cov}(x_{i,i-1}, x_{i+1, i})] = 2k^2(\sigma_x^2 - \text{Cov}(x_{i, i-1}, x_{i+1, i}))
\end{equation}
for each $i$.  Defining the spring-vector fluctuations $\delta x_{i+1,i} := x_{i+1,i} - \langle x_{i+1,i} \rangle$ then gives $\sum_i \delta x_{i+1,i} = 0$ and therefore 
\begin{align}
    0 = \left\langle \left(\sum_i \delta x_{i+1,i} \right)^2  \right\rangle %= \left\langle \sum_i^N \delta x_{i+1,i} \sum_j^N \delta x_{j+1,j}  \right\rangle 
    = \sum_i \langle \delta x_{i+1,i}^2 \rangle + \sum_{j, i \neq j} \langle \delta x_{i+1,i} \delta x_{j+1,j} \rangle = N\sigma_x^2+N(N-1)\text{Cov}(x_{i+1,i},x_{j+1,j}) .
\end{align}
It follows that $\text{Cov}(x_{i+1,i}, x_{j+1,j}) = -\sigma_x^2/(N-1)$ and therefore 
\begin{equation}
    \sigma_z^2 = 2k^2\left(\sigma_x^2+\frac{\sigma_x^2}{N-1}\right) = \frac{2k^2}{\beta k}\left(1-\frac{1}{N}\right)\left(1+\frac{1}{N-1}\right) = \frac{2k}{\beta} ,
\end{equation}
i.e. $z_i \sim \mathcal{N}(0, 2k/\beta)$. The mean event rate of non-factorised ECMC is therefore 
\begin{align}
    \Lambda_{\rm nf} &= \langle \lambda \rangle = \beta \langle \max(0, v z_i) \rangle = \frac{\beta}{2}\langle |v z_i| \rangle \nonumber = \frac{\beta|v|}{2}\int_{-\infty}^\infty |z|\frac{1}{\sqrt{2\pi \sigma_z^2}}\exp\left(\frac{-z^2}{2\sigma_z^2}\right)dz =\beta v_0 \int_{0}^\infty \frac{z}{\sqrt{2\pi \sigma_z^2}}\exp\left(\frac{-z^2}{2\sigma_z^2}\right)dz
    \\ &= \frac{\beta v_0}{\sqrt{2\pi\sigma_z^2}}\sigma_z^2=\frac{\beta v_0 \sigma_z}{\sqrt{2\pi}} = \frac{\beta v_0}{\sqrt{2\pi}}\sqrt{\frac{2k}{\beta}} = \sqrt{\frac{\beta k}{\pi}} v_0 ,
\end{align}
where we have used $\max(0, x) = (x + | x |) / 2$.

\subsubsection{Cell-horizon ECMC}

The cell-horizon gradients related to particle $i$ are
\begin{equation}
    \frac{\partial \widetilde{U}_{i, i-1}}{\partial x_i} = kx_{i,i-1}-b \quad \textrm{and} \quad \frac{\partial \widetilde{U}_{i+1, i}}{\partial x_i} = b - kx_{i+1,i} .
\end{equation}
%with $x_{i+1,i}\sim\mathcal{N}(L/N, (\beta k)^{-1}(1-N^{-1}))$. 
Recalling that $x_{i+1,i}\sim\mathcal{N}(L/N, (\beta k)^{-1}(1-N^{-1}))$ for each $i$ (see Section~\ref{app:SpringVecDistr}), the mean event rate of cell-horizon ECMC is therefore 
\begin{align}
    \Lambda_{\rm ch} &= \langle \widetilde{\lambda}_{i, i-1} \rangle + \langle \widetilde{\lambda}_{i+1, i} \rangle = \beta k\left[\left\langle \max\left(0, v\left(x_{i,i-1}-\frac{b}{k}\right)\right)\right\rangle + \left\langle \max\left(0, v\left(\frac{b}{k}-x_{i+1,i}\right)\right)\right\rangle\right] \nonumber \\
    &= \frac{\beta k|v|}{2}\left[ \left\langle \left|x_{i,i-1}-\frac{b}{k}\right| \right\rangle + \left\langle \left|x_{i+1,i}-\frac{b}{k}\right| \right\rangle \right] = \beta k v_0 \left\langle \left|x_{i+1,i}-\frac{b}{k}\right|\right\rangle \nonumber \\
    &= \beta k v_0 \int_{-\infty}^{\infty}\left|x-\frac{b}{k}\right|\frac{1}{\sqrt{2\pi\sigma_x^2}}\exp\left(-\frac{(x-\mu)^2}{2\sigma_x^2}\right)dx ,
\end{align}
where $\mu = \langle x_{i+1,i} \rangle$ and we have again used $\max(0, x) = (x + | x |) / 2$.  Now let $u = x - \mu$ and $\Delta = b/k - \mu$:
\begin{align}
    \Lambda_{\rm ch} &= \beta k v_0 \int_{-\infty}^{\infty} |u-\Delta| \frac{1}{\sqrt{2\pi\sigma_x^2}} \exp\left(-\frac{u^2}{2\sigma_x^2}\right) du \nonumber\\
    &= \beta k v_0 \left[ \int_{-\infty}^{\Delta} (\Delta-u) \frac{1}{\sqrt{2\pi\sigma_x^2}} \exp\left(-\frac{u^2}{2\sigma_x^2}\right) du + \int_{\Delta}^{\infty} (u-\Delta) \frac{1}{\sqrt{2\pi\sigma_x^2}} \exp\left(-\frac{u^2}{2\sigma_x^2}\right) du \right] \nonumber\\
    &= \beta k v_0 \left[\Delta \text{erf}\left(\frac{\Delta}{\sqrt{2}\sigma_x}\right) + \sqrt{\frac{2}{\pi}}\sigma_x\exp\left(\frac{-\Delta^2}{2\sigma_x^2}\right)\right]
\end{align}
We can now determine the optimal $\Delta$ by minimising $\Lambda_{\rm ch}$:
\begin{equation}
    \frac{d \Lambda_{\rm ch}}{d\Delta} = \beta k v_0 \left[ \text{erf}\left(\frac{\Delta}{\sqrt{2}\sigma_x}\right) + \frac{\Delta}{\sqrt{2}\sigma_x} \frac{2}{\sqrt{\pi}} \exp\left(\frac{-\Delta^2}{2\sigma_x^2}\right)  - \sqrt{\frac{2}{\pi}} \frac{\Delta}{\sigma_x} \exp\left(\frac{-\Delta^2}{2\sigma_x^2}\right) \right] = \beta k v_0 \, \text{erf}\left(\frac{\Delta}{\sqrt{2}\sigma_x}\right) ,
\end{equation}
which implies that $d\Lambda_{\rm ch} / d \Delta = 0$ when $\Delta=0$, i.e. $b^*_\Lambda = Lk/N$.  The mean event rate at $b=b^*_\Lambda$ is therefore 
\begin{equation}\label{eq:FactorisedMeanEventRate}
    \Lambda_{\rm ch}(\Delta=0) = \beta k v_0 \sqrt{\frac{2}{\pi}}\sigma_x = \sqrt{\frac{2 \beta k}{\pi}\left(1-\frac{1}{N}\right)} v_0 
\end{equation}
%where we have used Eq.~\eqref{eq:HarmChainVariance}.  
and it follows that
\begin{equation}
    \frac{\Lambda_{\rm ch}(\Delta=0)}{\Lambda_{\rm nf}} = \sqrt{2\left( 1 - \frac{1}{N} \right)} ,
\end{equation}
i.e. $\Lambda_{\rm ch}(\Delta=0) > \Lambda_{\rm nf}$ for all $N > 2$.  Moreover, the thermodynamic limit of Eq.~\eqref{eq:FactorisedMeanEventRate} is 
\begin{equation}
    \lim_{N\rightarrow \infty}\Lambda_{\rm ch}(\Delta = 0) = \sqrt{\frac{2\beta k}{\pi}} v_0
\end{equation}
and 
\begin{equation}
    \lim_{N\rightarrow \infty}\frac{\Lambda_{\rm ch}(\Delta=0)}{\Lambda_{\rm nf}} = \sqrt{2} .
\end{equation}

\subsection{Pressure}\label{harm_chain_appendix_pressure}

The partition function of the generalised harmonic chain (with total potential $\widetilde{U}$ defined in Eq.~\eqref{hc_factor1}) is 
\begin{equation}
    Z = \frac{1}{N!} \prod_{i=1}^N \left[ \int dx_{i+1,i} \right] \delta\left(\sum_i x_{i+1,i}-L \right)\exp\left(-\frac{\beta k}{p} \sum_i \left| x_{i+1,i} \right|^p +\beta bL\right) ,
\end{equation}
where the Dirac measure enforces the constraint $\sum_i x_{i+1,i}=L$.  Defining $y_i := x_{i+1,i} / L$ then gives 
\begin{equation}
    Z = L^{N-1} \frac{1}{N!} \prod_{i=1}^N \left[ \int dy_i \right] \, \delta \left( \sum_i y_i - 1 \right) \exp \left( -\frac{\beta k L^p}{p} \sum_i \left| y_i \right|^p +\beta bL \right) ,
\end{equation}
where the $L^{N-1}$ factor is the Jacobian when accounting for the Dirac measure.  The normalised pressure is then 
\begin{align}
    \beta P &= \frac{\partial \log Z}{\partial L}  \nonumber \\
    &= \frac{N-1}{L} + \frac{1}{Z} \left[ L^{N-1} \frac{1}{N!} \prod_{i=1}^N \left[ \int dy_i \right] \, \delta \left( \sum_i y_i - 1 \right) \left( -\beta k L^{p-1} \sum_i \left| y_i \right|^p + \beta b \right) \exp \left( -\frac{\beta k L^p}{p}\sum_i\left| y_i \right|^p +\beta bL \right) \right] \nonumber \\
    &= \frac{N-1}{L} + \frac{1}{Z} \left[ \frac{1}{N!} \prod_{i=1}^N \left[ \int dx_{i+1,i} \right] \, \delta \left( \sum_i x_{i+1,i} - L \right) \left( -\frac{\beta k}{L} \sum_i \left| x_{i+1,i} \right|^p + \beta b \right) \exp \left( -\frac{\beta k}{p}\sum_i\left| x_{i+1,i} \right|^p +\beta bL \right) \right] \nonumber \\
    &= \frac{N-1}{L} - \frac{\beta p}{L} \left\langle \frac{k}{p} \sum_i \left| x_{i+1,i} \right|^p \right\rangle + \beta b \label{eq:PressureGHC1} \\
    &= \frac{N-1}{L} - \frac{\beta p}{L} \left\langle \widetilde{U}(b=0) \right\rangle + \beta b . \label{eq:PressureGHC2}
\end{align}
This can be alternatively derived by writing the partition function in terms of an $(N-1)$-dimensional integral:
\begin{equation}
    Z = \frac{1}{N!} \prod_{i=1}^{N-1} \left[ \int dx_{i+1,i} \right] \exp\left(-\frac{\beta k}{p}\left[\sum_{i=1}^{N-1}\left|x_{i+1,i}\right|^p + \left| L-\sum_{i=1}^{N-1}x_{i+1,i} \right|^p\right] + \beta bL\right) .
\end{equation}
The change of variables then results in the same $L^{N-1}$ Jacobian.  We note that the pressure of the soft-disk model (Eq.~\eqref{eq:SoftDiskPressure}) can be analogously derived, resulting in the $- \beta p \left\langle \widetilde{U}(b=0) \right\rangle / L$ term in Eq.~\eqref{eq:PressureGHC2} becoming $\beta\kappa \left\langle \widetilde{U}(b=0) \right\rangle / L$, with $\widetilde{U}$ defined for the soft-disk model in Eq.~\eqref{eq:soft_disk_potential}.

For the $p=2$ case of the generalised harmonic chain, we use $\left\langle x_{i+1,i}^2 \right\rangle = \text{Var}(x_{i+1,i}) + \left\langle x_{i+1,i} \right\rangle^2 = (\beta k)^{-1}(1-1/N) + (L/N)^2$ (by Eq.~\eqref{eq:HarmChainVariance}) to simplify Eq.~\eqref{eq:PressureGHC1} to Eq.~\eqref{eq:HarmChainPressure}:
\begin{align}
    \beta P &= \frac{N-1}{L} - \frac{\beta k}{L} \left\langle \sum_i x_{i+1,i}^2 \right\rangle + \beta b \nonumber\\
    &= \frac{N-1}{L} - \frac{N}{L}\beta k \left\langle x_{i+1,i}^2 \right\rangle + \beta b \nonumber\\
    &= \frac{N-1}{L} - \frac{N}{L} \left( 1 - \frac{1}{N} \right) - \frac{L}{N}\beta k + \beta b \nonumber\\
    &= \beta \left( b - \frac{Lk}{N} \right) .
\end{align}

\begin{figure}[t!]
    \centering
    % Left Figure
    \begin{minipage}[t]{0.48\linewidth}
        \centering
        \includegraphics[width=\linewidth]{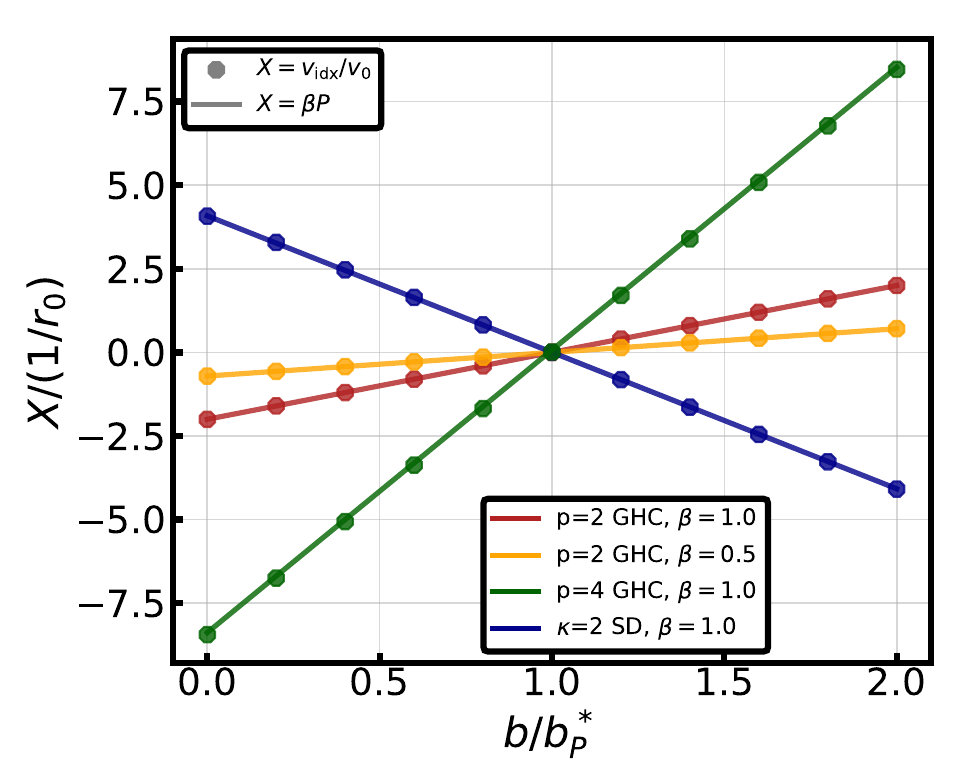}
        \caption{MC estimates of both $\beta P$ (lines; defined in Section~\ref{sec:FactorFieldECMC}) and $v_{\rm idx} / v_0$ (circles; defined in Section~\ref{sec:HarmonicChain}) vs $b$ for the generalised harmonic chain (GHC) and 1D soft-disk model (SD) with a variety of model exponents and inverse temperatures. Results are consistent with the ansatz $\beta P = v_{\rm idx} / v_0$.  Lines are a guide to the eye between pressure estimates computed at the same $b$ values as the $v_{\rm idx}$ estimates.  The length scale $r_0 = (\beta k)^{-1/p}$ and $\sigma$ for the GHC and SD model, respectively.  Results were averaged across 10 independent simulations, each of between $4.5 \times 10^5$ and $7 \times 10^5$ samples (see Appendix~\ref{sec:CodeAndSimulationData} for precise details).}
        \label{fig:index_velocity_pressure}
    \end{minipage}
    \hfill % Adds horizontal spacing
    % Right Figure
    \begin{minipage}[t]{0.48\linewidth}
        \centering
        \includegraphics[width=\linewidth]{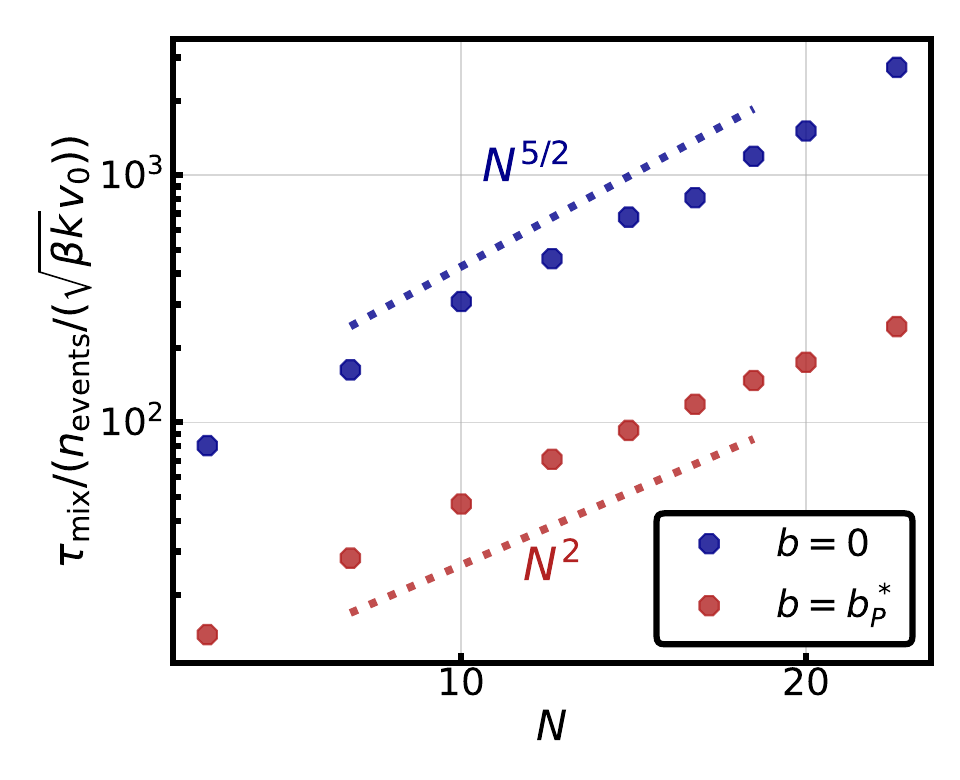}
        \caption{Estimates of the mixing times $\tau_{\rm mix}$ of standard-factorised ECMC (blue) and optimal cell-horizon ECMC (green) vs $N$ for the $p=2$ generalised harmonic chain. Mixing times are averaged across 12 independent simulations. The mixing-time estimate of any given simulation is defined as the time at which the instantaneous variance ${\rm var}(\mathbf{w})$ (defined in Appendix~\ref{app:Vidx}) reaches its equilibrium value after starting far from equilibrium. Results are consistent with standard-factorised ECMC and optimal cell-horizon ECMC achieving $\mathcal{O}(N^{5/2})$ and $\mathcal{O}(N^2)$ mixing times. Dotted lines are straight lines with gradient $N^{5/2}$ (blue) and $N^2$ (green).  %\blue{***$y$ axis: $\tau_{\rm mix} / (n_{\rm events} \tau_{\rm tot} / N) \to \tau_{\rm mix} / (n_{\rm events} / (\sqrt{\beta k} v_0))$***}
        }
        \label{fig:MixingTimes}
    \end{minipage}
\end{figure}

%\section{Index-space velocity and pressure}
\section{Additional numerical results}
\label{app:Vidx}

%This appendix presents numerical results to support the ansatz $\beta P  = v_{\rm idx}$.  

Here we present additional numerical results to support the ansatz $\beta P  = v_{\rm idx} / v_0$ and the mixing times of
Section~\ref{sec:Mixing}.  Fig.~\ref{fig:index_velocity_pressure} presents estimates of both $\beta P$ and $v_{\rm idx} / v_0$ for all models presented in the paper.  Results demonstrate excellent agreement between the two quantities.
 
Fig.~\ref{fig:MixingTimes} presents estimates of the mixing times $\tau_{\rm mix}$ of both standard-factorised ECMC and optimal cell-horizon ECMC.  Mixing times are averaged over $12$ independent simulations.  The mixing-time estimate of any given simulation is defined as the time at which the instantaneous variance ${\rm var}(\mathbf{w}) := \sum_i w_i^2 / N - \left( \sum_i w_i / N\right)^2$ of the \emph{half-system distances} $w_i := x_{i+N/2} - x_i$ reaches its equilibrium value from the far-from-equilibrium initial configurations described in Section~\ref{sec:Mixing} ($x_{i+N/2}$ accounts for the toroidal topology of the index space $\mathcal{I}$).  This is the method presented in Ref.~\cite{Lei2019ECMCwithFactorFields}.  Results are consistent with standard-factorised ECMC and optimal cell-horizon ECMC achieving $\mathcal{O}(N^{5/2})$ and $\mathcal{O}(N^2)$ mixing times for the $p=2$ generalised harmonic chain.

\twocolumngrid

\bibliographystyle{ieeetr}
\bibliography{ref,Faulkner}

\end{document}